\documentclass[aps,epsfig,graphics,floatfix,twocolumn,mathbbm,nopacs,nofootinbib]{revtex4}

\usepackage{amsmath,amsfonts,amssymb,amsthm,stmaryrd,graphics,graphicx,epsfig,color,times,bbm,psfrag}

\usepackage[all]{xy}

\newcommand{\xnode}[1]{*+[F]{#1}}
\newcommand{\xmnode}[1]{*+[F]{A[#1]}}
\newcommand{\xminid}[1]{
	\xymatrix@C=5mm{#1}
}
\newcommand{\Par}{\ar@{-}[r]}
\newcommand{\Pau}{\ar@{-}[u]}
\newcommand{\Pad}{\ar@{-}[d]}
\newcommand{\Pal}{\ar@{-}[l]}

\begin{document}

\bibliographystyle{apsrev}

\newcommand{\tr}{\operatorname{tr}}
\newcommand{\uinvnorm}{|\kern-2pt|\kern-2pt|}
\newcommand{\wt}{\operatorname{wt}}
\newcommand{\spectrum}{\operatorname{sp}}
\newcommand{\erf}{\operatorname{erf}}
\newcommand{\erfc}{\operatorname{erfc}}
\newcommand{\supp}{\operatorname{supp}}
\newcommand{\diag}{\operatorname{diag}}

\bibliographystyle{apsrev}

\newcommand{\me}{\mathrm{e}}
\newcommand{\mi}{\mathrm{i}}
\newcommand{\md}{\mathrm{d}}

\newcommand{\cc}{\mathbb{C}}
\newcommand{\nn}{\mathbb{N}}
\newcommand{\rr}{\mathbb{R}}
\newcommand{\zz}{\mathbb{Z}}

\newcommand{\id}{\mathbbm{1}}

\newtheorem{theorem}{Theorem}
\newtheorem{definition}[theorem]{Definition}
\newtheorem{lemma}[theorem]{Lemma}
\newtheorem{corollary}[theorem]{Corollary}
\newtheorem{property}[theorem]{Property}
\newtheorem{proposition}[theorem]{Proposition}
\newtheorem{remark}[theorem]{Remark}
\newtheorem{example}[theorem]{Example}
\newtheorem{assumption}[theorem]{Assumption}
\newtheorem{observation}[theorem]{Observation}

\setlength{\parskip}{2pt}

\newcommand{\identity}{\id}
\newcommand{\bra}[1]{\mbox{$\langle #1 |$}}
\newcommand{\ket}[1]{\mbox{$| #1 \rangle$}}
\newcommand{\braket}[2]{\mbox{$\langle #1  | #2 \rangle$}}
\newcommand{\proj}[1]{\mbox{$|#1\rangle \!\langle #1 |$}}
\newcommand{\ev}[1]{\mbox{$\langle #1 \rangle$}}
\def\sign{\mbox{sgn}}
\def\H{{\cal H}}
\def\C{{\cal C}}
\def\E{{\cal E}}
\def\O{{\cal O}}
\def\B{{\cal B}}
\def\one{\ensuremath{\hbox{$\mathrm I$\kern-.6em$\mathrm 1$}}}
%{\ensuremath{\hbox{$\mathrm I$\kern-.6em$\mit 1$}}}
\def\tr{ \mbox{tr}}

\newcommand{\note}[1]{{\sc #1}}

\bibliographystyle{unsrt}

\title{Measurement-based quantum computation beyond the one-way model}

\author{D.\ Gross and J.\ Eisert}

\affiliation{
Blackett Laboratory, 
Imperial College London,
Prince Consort Road, London SW7 2BW, UK\\
Institute for Mathematical Sciences, Imperial College London,
Exhibition Rd, London SW7 2BW, UK}

\author{N.\ Schuch}

\affiliation{Max-Planck-Institut f{\"u}r Quantenoptik, 
Hans-Kopfermann-Str.\ 1, 85748 Garching, Germany}

\author{D.\ Perez-Garcia}

\affiliation{Departamento de Analisis Matematico, Universidad Complutense de Madrid,
28040 Madrid, Spain}

\date\today

\begin{abstract}
We introduce novel schemes for 
quantum computing based on local measurements on 
entangled resource states. 
This work elaborates on the framework established 
in [Phys.\ Rev.\ Lett.\ {\bf 98}, 220503  (2007), quant-ph/0609149]. 
Our method 
makes use of tools from many-body
physics -- matrix product states, finitely correlated states or
projected entangled pairs states --  to show how measurements on
entangled states can be viewed as processing quantum information. 
This work hence constitutes an instance where a
quantum information problem -- how to realize
quantum computation -- was approached using tools
from many-body theory and not vice versa. 
We give a more detailed description of the setting, and 
present a large
number of new examples.  We find novel computational 
schemes, which differ from the original one-way computer for 
example in the way the
randomness of measurement outcomes is handled. Also, schemes are
presented where the logical qubits are no longer strictly localized on
the resource state. 
Notably, we find a great flexibility in the properties of the
universal resource states: They may for example exhibit non-vanishing
long-range correlation functions or be locally arbitrarily close to a
pure state.
We discuss variants of Kitaev's toric code states as universal
resources, and contrast this with situations where they can be
efficiently classically simulated.  This framework opens up a way of
thinking of tailoring resource states to specific physical systems,
such as cold atoms in optical lattices or linear optical systems.
\end{abstract}


\maketitle

\section{Introduction}

Consider a quantum state of some system consisting of many particles.
This system could be a collection of cold atoms in an optical lattice,
or of atoms in cavities, coupled by light, or entirely optical
systems. Assume that one is capable of performing local projective
measurements on that system, however there is no way to realize a
controlled coherent evolution.  Can one perform universal quantum
computing in such a setting?  Perhaps surprisingly, this is indeed the
case: The  {\it one-way model} of Refs.\ \cite{Oneway,OnewayCompModel}
demonstrates that local measurements on the {\it cluster state} -- 
a certain multi-particle entangled state on an array of qubits
\cite{Cluster} -- do
possess this computational power.  The insight gives rise to an
appealing view of quantum computation: One can in principle 
abandon
the need for any unitary control, once the initial state has been
prepared.  The local measurements -- a feature that any computing
scheme would eventually embody -- then take the role of preparation of
the input, the computation proper, and the read-out.  This is of
course a very desirable feature: Quantum computation then only amounts to (i) preparing a universal resource state and (ii) 
performing local
projective measurements [2--6]. 

But what about other entangled quantum states, different from
cluster or graph states \cite{GS,Survey}? Can they form a resource for universal computation?  Is it
possible to tailor resource states to specific physical systems?  For
some experimental implementations -- e.g.,\ cold atoms in optical
lattices \cite{Greiner}, atoms in cavities \cite{Plenio,Atoms}, optical
systems [11-13], 
ions in traps \cite{Traps}, or many-body ground
states -- it may well be that preparation of cluster states is
unfeasible, costly, or that they are particularly fragile to finite
temperature or decoherence effects. Also, from a fundamental point of
view, it is clearly interesting to investigate the computational power
of many-body states -- either for the purpose of building
measurement-based quantum computers or else for deciding which states
could possibly be classically simulated \cite{vidal,maarten2}. Interestingly,
very little progress has been made over the last years when it comes
to going beyond the cluster state as a resource for measurement-based
quantum computation (MBQC). To our knowledge, 
no single computational
model distinct from the one-way computer has been developed which
would be based on local measurements on an algorithm-independent qubit
resource state. 

The apparent lack of new schemes for MBQC is all the 
more surprising, given the great advances that have been 
made toward an understanding
of the structure of cluster state-based computing itself. 
For example, it has
been shown that the computational model of the 
one-way computer and
teleportation-based approaches to quantum 
computing \cite{teleporters} are essentially 
equivalent \cite{teleportersEquiv,Jozsa}. 
A particularly elegant way of realizing
this equivalence was discovered in Ref.\ 
\cite{OneWayPEPS}: They pointed out that the maximally entangled
states used for the teleportation need not be physical. Instead, the
role can be taken on by virtual entangled pairs used in a ``valence
bond'' \cite{aklt} description of the cluster state. This point of
view is closely related to our approach to be described below. 
Further progress includes a clarification of the 
temporal inter-dependence of measurements \cite{elham}.
In Ref. \cite{Vedral} a first non-cluster (though not universal, but
algorithm-dependent) resource has been introduced, 
which includes the natural ability of
performing three-qubit gates. Recently, Refs.\ 
\cite{maarten1,InnsbruckLong} 
initiated a detailed study of resource 
states which can be used to prepare cluster states 
(see Section \ref{sec:propertiesDiscussion}).

In this work, we describe methods for the systematic construction of
new MBQC schemes and resource states. 
This continues a program
initiated in Ref.\ \cite{shortOne} in a more detailed fashion.  We
analyze MBQC in terms of ``computational tensor networks'', building
on a familiar tool from many-body physics known by the names of
matrix-product states, finitely correlated states \cite{FCS,david} or
projected entangled pair states \cite{PEPS,aklt}. 

The problem of finding novel schemes for measurement-based computation can be approached from two different points of 
view. Firstly, one may
concentrate on the \emph{quantum states} which provide the
computational power of measurement-based computing schemes and ask
\begin{enumerate}
	\item
	{\it What are the properties that render a state a universal resource for a
	measurement-based computing 
	scheme\footnote{
		Clearly, the answer to the previous question depends on the
		definition of a \emph{universal resource}. See Section
		\ref{sec:propertiesDiscussion} for a discussion, in particular in
		relation to Ref.\ \cite{maarten1}.
	}?	
	}
\end{enumerate}
Secondly, putting the emphasize on
\emph{methods}, the central question becomes
\begin{enumerate}
	\setcounter{enumi}{1}
	\item {\it 
	How can we systematically construct new schemes for measurement-based
	quantum computation?
	Is there a framework which is flexible enough to allow for the
	construction of a variety of different models?}
\end{enumerate}
Both of these intertwined questions will be addressed in this work.

\section{Main results}

As our main result, we present a plethora of new universal resource
states and computational schemes for MBQC. The examples have been
chosen to demonstrate the flexibility one has when constructing models
for measurement-based computation. Indeed, it turns out that many
properties one might naturally conjecture to be necessary for a state
to be a universal resource can in fact be relaxed.  Needless to say,
the weaker the requirements are for a many-body state to form a
resource for quantum computing, the more feasible physical
implementations of MBQC become.

Below, we enumerate some specific results concerning the properties of
resource states. The list pertains to Question 1 given in the
introduction.
\begin{itemize}
	\item
	In the cluster state, every particle is maximally entangled with the
	rest of the lattice. Also, the localizable entanglement \cite{LE} is
	maximal (i.e.\ one can deterministically prepare an maximally
	entangled state between any two sites, by performing local
	measurements on the remainder). While both properties are essential
	for the original one-way computer, they turn out not to be necessary
	for computationally universal resource states. To the contrary, we
	construct \emph{universal states which are locally arbitrarily pure}. 

	\item
	For previously known schemes for MBQC,
	it was essential that far-apart
	regions of the state were uncorrelated. This feature allowed one to
	logically break down a measurement-based calculation into small
	parts corresponding to individual quantum gates. Our framework does
	not depend on this restriction and resources with {\it non-vanishing
	correlations} between any two subsystems are shown to exist.
	This property is common e.g.,\ in many-body ground-states.
	%In fact, we will see that 
	%non-vanishing correlations throughout the lattice -- say, 
	%exponentially decaying correlation functions, as common in gapped local
	%models  --  are no obstacle to a state being a universal resource.

	\item 
	Cluster states can be prepared step-wise by means of a bi-partite
	\emph{entangling gate} (controlled-phase gate). 
	%Technically
	%speaking, the cluster results from the action of a 
	%quantum cellular
	%automaton (QCA) \cite{QCA}. 
	This property
	is important to the original universality proof. More generally, one
	might conjecture that resource states must always result from an
	entangling process making use of mutually commuting entangling
	gates, also known as a unitary 
	\emph{quantum cellular automaton}
	\cite{QCA}. Once more, this requirement turns out not to be
	necessary.

	\item
	The cluster states can be used as \emph{universal preparators}: Any
	quantum state can be distilled out of a sufficiently large cluster
	state by local measurements. Once more, this property is essential
	to the original one-way computer scheme. However, computationally
	universal resource states not exhibiting this properties do exist
	(the reader is referred to Ref.\ \cite{maarten1} for an analysis of
	resource states which are required to be preparators; see also the
	discussion in Section \ref{sec:propertiesDiscussion}).  More
	strongly, we construct universal resources out of which not even a
	single two-qubit maximally entangled state can be distilled.

	\item
	 A genuine \emph{qu-trit} resource is presented (distinct, of
	 course, from a qu-trit version of the cluster state 
	 \cite{dLevel}).
\end{itemize}

\begin{figure}
	\includegraphics[width=5.5cm]{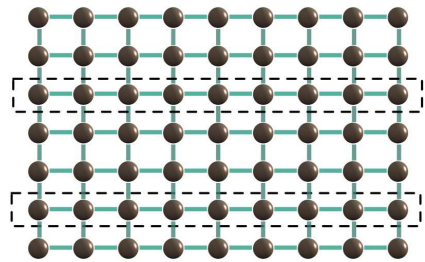}
	\caption{\label{fig:flow}
		Measurement-based quantum computing as generalization of the
		one-way model as being considered in this work. Initially, an
		entangled resource state is available, different from 
		the cluster state, followed by local projective 
		measurements on all individual constituents in the regular
		not necessarily cubic lattice. In all figures, dark gray 
		circles denote individual physical systems.}
\end{figure}

%Further, we will argue in Section \ref{sec:propertiesDiscussion} that
%giving an algorithm for efficiently classically simulating any
%measurement scheme on a state is currently the only way to disprove
%its potential usefulness as a universal computational resource.

We will further see that there is quite some flexibility concerning the
computational model itself (addressing Question~2 mentioned in the
introduction):
\begin{itemize}
\item	
	The new schemes differ from the one-way model in the way the
	\emph{inherent randomness} of quantum measurements is dealt with.

\item
	We generalize the well-known concept of \emph{by-product operators}
	to encompass any finite group. E.g.\ we show the existence of
	computational models, where the by-product operators are elements of
	the entire single-qubit Clifford group, or the dihedral group.

\item 
	We explore schemes where each logical qubit is encoded in \emph{several
	neighboring correlation systems} 
	(see Section \ref{sec:ctn} for a
	definition of the term ``correlation system'').
	
\item 
	One can find ways to construct schemes in which interactions between
	logical qubits are controlled by ``routing'' the qubits towards an
	``interaction zone'' or keeping them away from it.

\item 
	In many schemes, we adjust the layout of the measurement pattern
	dynamically, incorporating information about previous measurement
	outcomes as we go along. In particular, the expected length of a computation is
	random (this constitutes no problem, as the probability of exceeding
	a finite expected length is exponentially small in the excess).
\end{itemize}

\subsection{Universal resource states}
\label{sec:propertiesDiscussion}

What are the properties from which a universal resource state derives
its power? After clarifying the terminology, we will argue that an
answer to this question -- desirable as it may be -- faces formidable
obstacles.

Quantum computation can come in a variety of different 
incarnations, as diverse as e.g.,\ the well-known gate-model 
\cite{nielsenChuang},
adiabatic quantum computation \cite{adiabatic} or MBQC. All these
models turn out to be equivalent in that they can simulate each other
efficiently. 

For measurement-based schemes, the ``hardware'' consists of a
multi-particle quantum system in an algorithm-independent state and a
classical computer. The input is a gate-model description of a quantum
computation. In every step of the computation, a local measurement is
performed on the quantum state and the result is fed into the
classical computer. Based on the outcomes of previous steps, the
computer calculates which basis to use for the next measurements and,
finally, infers the result of the computation from the measurement
outcomes \cite{OnewayCompModel}. Having this procedure in mind, we
call a quantum state a \emph{universal resource} for MBQC, if a
classical computer assisted by local measurements on this states can
efficiently predict the outcome of any quantum computation.

The reader should be aware that another approach has recently
been described in
the literature. The cluster state has actually a stronger property
than the one just used for the definition of universality: it is a
universal preparator. This means that one can prepare any given
quantum state on a given sub-set of sites of a sufficiently large
cluster by means of local measurements. Hence, cluster states could in
principle be
used for information processing tasks which require a quantum output.
Ref.\ \cite{InnsbruckLong} referred to this scenario as
{\it CQ-universality} -- i.e.\ universality for problems which require a
classical input but deliver a quantum output. This observation is the
basis of Ref.\ \cite{maarten1}, where a state is called a universal
resource if it possesses the strong property of being a universal
preparator, or, equivalently, of being CQ-universal. 

Clearly, any efficient universal preparator is also a computationally universal
resource for MBQC (since one can, in particular, prepare the cluster
state). But the converse is not true, as our results show. Indeed,
while it proves possible to come up with necessary criteria for a
state to be a universal preparator \cite{maarten1}, we will argue
below that the current limited understanding of quantum computers
makes it extremely hard to specify necessary conditions for
computational universality.

In order to pinpoint the source of the quantum speedup, we might try
to find schemes where more and more work is done by the classical
computer, while the employed quantum states become ``simpler'' (e.g.,\
smaller or less entangled). How far can we push this program without
losing universality? The answer is likely to be intractable.
Currently, we are not aware of a proof that quantum computation is
indeed more powerful than classical methods. Hence, it can
presently not be excluded that no assistance from a quantum state is 
necessary at all.

\begin{observation}[Any state may be a universal resource]\label{obs:assumption}
	If one is unwilling to \emph{assume} that there is a 
	separation between classical and quantum computation (i.e., BPP $\neq$ BQP), 
	then it is
	impossible to rule out any state as a universal resource.
\end{observation}

It is, however, both common and sensible to assume superiority of
quantum computers and we will from now on do so. Observation
\ref{obs:assumption} still serves a purpose: it teaches us that the
only known way to rule out universality is to invoke this 
assumption (this 
avenue was taken, e.g.,\ in Refs.\ \cite{bravyi:toric-mbc,maarten2}).

\begin{observation}[Efficient classical simulation]\label{obs:onlyway}
	The only currently known method for excluding the possibility that a
	given quantum state forms a universal resource is to show that any
	measurement-based scheme utilizing the state can be efficiently
	simulated by a classical computer.  
\end{observation}

Thus, the situation presents itself as follows: 
there is a tiny set of quantum states for which it is possible to
prove that any local measurement-based scheme can be efficiently simulated. On the
other extreme, there is an even tinier set for which universality is
provable. For the vast majority no assessment can be made.
Furthermore, given the fact that rigorously establishing the
``hardness'' of many important problems in computer science turned out
to be extremely challenging, it seems unlikely that this
situation will change dramatically in the foreseeable future.

We conclude that a search for necessary conditions for universality is
likely to remain futile. The converse question, however, can be
pursued: it is possible to show that many properties that one might
naively assume to be present in any universal resource are, in fact,
unnecessary.

\section{Computational tensor networks}
\label{sec:ctn}

The current section is devoted to an in-depth treatment of a class of
states known respectively as valence-bond states, finitely correlated
states, matrix product states or projected entangled pairs states,
adapted to our purposes of measurement-based quantum computing. This
family turns out to be especially well-suited for a description of a
computing scheme. 

Indeed, any systematic analysis of resources states requires a
framework for describing quantum states on extended systems. We
briefly compile a list of desiderata, based on which candidate
techniques can be assessed.

\begin{itemize}
	\item
	The description should be \emph{scalable}, so that a class of states
	on systems of arbitrary size can be treated efficiently.

	\item
	As quantum states which are naturally described in terms of
	one-dimensional topologies have been shown to be classically
	simulable \cite{FCS,vidal,maarten2}, the framework ought to handle {\it two- or higher dimensional
	topologies} naturally.

	\item
	The basic operation in measurement-based computation are {\it local
	measurements}.  It would be desirable to describe the effect of local
	measurements in a local manner.  Ideally, the class of efficiently
	describable states should be closed under local measurements.

	\item
	The class of describable states should include elements which show
	features that naturally occur in {\it ground states} of quantum
	many-body systems, such as {\it non-maximal local entropy of
	entanglement} or {\it non-vanishing two-point correlations}, etc.
\end{itemize}

The description of states to be introduced below complies with all of
these points.

We will introduce the construction in several steps, starting with
one-dimensional matrix product states. The new view on the
processing of information is that the matrices appearing
in the description of resource states are taken literally, as
operators processing quantum information.

\subsection{Matrix product states}

A {\it matrix product state} (MPS) 
for a chain of $n$ systems of physical dimension $d$
(so $d=2$ for qubits) 
is specified by
\begin{itemize}	
	\item
	An {\it auxiliary $D$ dimensional vector space} 
	($D$ being some parameter,
	describing the amount of correlation between two consecutive blocks
	of the chain),

	\item
	For each system $i$ a set of $d$ $D\times D$-{\it  matrices} $A_i[j],
	j\in\{0\dots d-1\}$.

	\item
	Two $D$-dimensional vectors $\ket L, \ket R$ representing {\it  
	boundary
	conditions}.
\end{itemize}

The state vector $\ket\Psi$ of the matrix product state
is then given explicitly by 
\footnote{
There is a reason why the \emph{right}-hand-side boundary condition
$\ket R$ appears on the \emph{left} of Eq.\ (\ref{eqn:linearMps}). In
linear algebra formulas, information usually flows from right to left:
$B A\ket\psi$ means ``$\ket\psi$ is acted on by $A$, then by $B$''.
In the graphical notation to be introduce later, it is much more
natural to let information flow from left to right:
\begin{equation}
	\xminid{\xnode{\ket\psi}\ar[r]&\xnode{A}\ar[r]&\xnode{B}\ar[r]&}.
\end{equation}
The order in Eq.\  (\ref{eqn:linearMps}) anticipates the graphical
notation.}
\begin{equation}\label{eqn:linearMps}
	 \ket\Psi=\sum_{s_1,\dots, s_n=0}^{d-1}
 	\bra R A_n[s_n] \dots A_1[s_1] \ket L \,\,\ket{s_1, \dots, s_n}. 
\end{equation}
From now on we will assume that the matrices are site-independent:
$A_i[j]=A[j]$, so the MPS is translationally invariant up
to the boundary conditions. We take the freedom of disregarding
normalization whenever this consistently possible.

Let us spend a minute interpreting Eq.\ (\ref{eqn:linearMps}). Assume
we have measured the first site in the computational basis and
obtained the outcome $s_1$. One immediately sees that the resulting
state vector $\ket{\Psi'(s_1)}$ on the remaining sites is again a MPS, 
where the left-hand side boundary vector now reads
\begin{equation}\label{eqn:firstComp}
	\ket{L'(s_1)}=A[s_1]\ket L. 
\end{equation}
Hence the state of the auxiliary system
gets changed according to the measurement outcome. So we find that the
correlations between the state of the first site and the rest of the
chain are mediated via the auxiliary space, which will thus be
referred to as \emph{correlation space} in the sequel.

In the past, the matrices appearing in the definition of $\ket\Psi$
have been treated mainly as a collection of variational parameters,
used to parametrize ansatz states for ground states of spin chains
\cite{FCS}. However -- and that is the basic insight underlying our
view on MBQC -- Eq.\ (\ref{eqn:firstComp}) can also be read as an
operator $A[s_1]$ acting on some quantum state $\ket L$. We will
elaborate on this interpretation in Section \ref{1DTN}.

In order to translate Eq.\ (\ref{eqn:linearMps}) to the setting of 2-D
lattices, we need to cast it into the form of a tensor network.
Setting $L_i=\braket iL $ and
\begin{equation}	
	A[s]_{i,j}:=\bra j A \ket i,
\end{equation}	
we can write Eq.\ 
(\ref{eqn:linearMps}) as
\begin{equation}\label{eqn:tensorMps}
	\langle s_1, \dots , s_n | \Psi \rangle = 
	\sum_{i_0,\dots,i_n}^D
	L_{i_0} 
	A[s_1]_{i_0,i_1} \dots A[s_n]_{i_{n-1}, i_n} 
	{R^\dagger}_{i_n}.
\end{equation}

While Eq.\ (\ref{eqn:tensorMps}) is 
awkward enough, the 2-D equivalent
is completely unintelligible. To cure this problem, we introduce a
graphical notation\footnote{These graphical formulae	
	are compatible with various similar systems introduced before
	\cite{graphical}.}
which enables an intuitive 
understanding beyond the 1-D case. In the
following, tensors will be represented by boxes, indices by
edges:
\begin{eqnarray}
		L_r&=&\xymatrix@C=5mm{*+[F]{L}\ar[r]&},\\
	  A[s]_{l,r}&=&\xymatrix@C=5mm{\ar[r]&*+[F]{A[s]}\ar[r]&},\\
	  {R^\dagger}_l &=&\xymatrix@C=5mm{\ar[r]&*+[F]{R^\dagger}}\,.
\end{eqnarray}
Needless to say, in the equation above, ``$l$'' is the index leaving
the box on the left-hand-side, ``$r$'' the right-hand-side one.
Connected lines designate contractions of the respective indices.
Eq.\ (\ref{eqn:linearMps}) now reads
\begin{equation*}
	\langle s_1, \dots , s_n | \Psi \rangle = 
	\xymatrix@C=5mm{
		\xnode{L}\ar@{-}[r]&*+[F]{A[s_1]}\ar@{-}[r]&
		\dots\ar@{-}[r]&*+[F]{A[s_n]}\ar@{-}[r]&*+[F]{R^\dagger}
	}.
\end{equation*}
A single-index tensor can be interpreted as the expansion coefficients
of either a ``ket'' or a ``bra''. Sometimes, we will indicate what
interpretation we have in mind by placing arrows on the edges:
outgoing arrows designating ``kets'', incoming arrows ``bras''
\begin{equation}
		\xymatrix@C=5mm{*+[F]{L}\ar[r]&}= \ket L,\quad
	  \xymatrix@C=5mm{\ar[r]&*+[F]{R^\dagger}}= \bra R.
\end{equation}
Tensors with two indices $A_{l,r}$ can naturally be interpreted as
operators. In the graphical notation we often want to think of
information flowing from the left to the right, in which case
$A=\sum_{l,r} A_{l,r} \ket r_r \bra l_l$ would be denoted as
\begin{equation}
	\xymatrix@C=5mm{\ar[r]&*+[F]{A}\ar[r]&}= A,
\end{equation}
i.e.\ with the l.h.s.\ index being associated with a ``bra'' and the
r.h.s\ one with a ``ket''. 
%\begin{eqnarray}
%	A=\sum_{l,r} A_{l,r} \ket r \bra l
%	=\sum_{l,r} A_{l,r} 
%	\xymatrix{&\ar[r]&*+[F]{l}&*+[F]{r}\ar[r]&}
%	=
%	A_{l,r}\xymatrix@C=5mm{\ar[r]&*+[F]{A[s]}\ar[r]&}.
%\end{eqnarray}
The following relations exemplify the definition:
\begin{eqnarray}
	\braket RL
	&=& \xminid{\xnode{L}\ar@{-}[r]&\xnode{R}}\,,\\
	\nonumber \\
	A \ket L
	&=& \xminid{\xnode{L} \ar@{-}[r]&\xnode{A} \ar[r]&},\\
	\nonumber \\
	A B&=&\xminid{\ar[r]&\xnode{B} \ar@{-}[r]&\xnode{A}\ar[r]&} ,\\
	\nonumber \\
	\tr (A B)&=& 
	\begin{xy}
		*!C\xybox{\xymatrix@C=5mm@R=3mm{
			&\xnode{B} \ar@{-}[r]& \xnode{A}\ar@{-} `/4pt[r]`[dl]`[dll]`[l][l] &   &\\
			&        &             &	 &
		}}
	\end{xy}.
\end{eqnarray}
The formula for the expansion coefficients of a matrix product state
finally becomes
\begin{equation*}
	\langle s_1, \dots , s_n | \Psi \rangle = \label{eq:ctn}\
	\xymatrix@C=5mm{
		\xnode{L} \ar@{-}[r]&*+[F]{A[s_1]} \ar@{-}[r]&
		\dots \ar@{-}[r]&*+[F]{A[s_n]} \ar@{-}[r]&*+[F]{R^\dagger}
	}\,.
\end{equation*}
This formula suggest a more ``dynamic'' interpretation of MPS: the
l.h.s.\ boundary conditions $\ket L$ specify an initial state of the
correlation system, which is acted on by the matrices of the MPS
representation. The next paragraph is going to elaborate on this
point.

\subsection{Quantum computing in correlation systems}\label{1DTN}

We return to the discussion of the properties of matrix product
states. Above, it has been shown how to compute the overlap of $\ket\Psi$
with an element of the computational basis (c.f.\ Eq.\
(\ref{eqn:tensorMps})). The next step is to
generalize this to any local projection operator.  Indeed, if
$\ket\phi$ is a general state vector in $\cc^2$, we abbreviate
\begin{equation}
	\braket\phi0\,A[0]+\braket\phi1\,A[1]=:A[\phi].  
\end{equation}
One then easily
derives the following, central formula
\begin{equation}\label{eqn:transport}
	\big(\bigotimes_i^n \bra{\phi_i}\big) \ket\Psi 
	= 
	\xymatrix@C=1mm@!{
		*+[F]{L}\Par[r]&*+[F]{A[\phi_1]}
		\Par[r]&\dots\Par&*+[F]{A[\phi_n]}\Par&*+[F]{R} 
	}\,\, .
\end{equation}

Now suppose we measure local observables on $\ket\Psi$ and obtain
results corresponding to the eigenvector $\ket{\phi_i}$ at the $i$-th
site. Eq.\ (\ref{eqn:transport}) allows us to re-interpret this
process as follows. Initially, the $D$-dimensional correlation system 
is prepared in the state $\ket L$. The result $\ket{\phi_1}$ at the
first site induces the evolution 
\begin{equation}
	\ket L \mapsto A[\phi_1]\ket L.
\end{equation}	
From this point of view, a sequence of measurements on $\ket\Psi$ is
tantamount to a processing of the correlation system's state by
the operations $A[\phi_i]$.\footnote{
	Of course, for general measurement bases, $A[\phi_i]$ is not going to
	be unitary. Choosing the bases in such a way as to ensure unitarity is
	an essential part of the design of a computational scheme for a given
	resource.
} 
An appealing perspective on MBC suggests
itself:

\begin{observation}[Role of correlation space]
Measurement-based computing takes place in
correlation space. The gates acting on the correlation systems are
determined by local measurements.  Intuitively, ``quantum
correlations'' are the source of a resource's computational potency.
The strength of this framework lies in the fact that it assigns a
concrete mathematical object to these correlations.
\end{observation} 

Indeed, it will turn out that MBQC can be understood completely using
this interpretation. 

\subsection{Example: The 1-D cluster state}
\label{sec:1dcluster}

To illustrate the abstract definitions made above, we will discuss the
linear cluster state vector $\ket{Cl_n}$ in this section. It is both one of
the simplest and certainly the most important MPS in the context of
MBQC.

What is the tensor network representation of $\ket{Cl_n}$?
Recall that the cluster state can be generated by preparing $n$ sites
in the state vector $\ket+:= \ket0+\ket1$ and subsequently applying
the controlled-$Z$ operation
\begin{equation}
	CZ = |0,0\rangle\langle 0,0 | + |0,1\rangle\langle 0,1|+
	|1,0\rangle\langle 1,0|  - |1,1\rangle\langle 1,1|
\end{equation}
between any two nearest neighbors. Effectively, $CZ$ introduces a
$\pi$-phase whenever two consecutive systems are in the $\ket1$-state.
Hence its expansion coefficients in the computational basis are given
by
\begin{equation}\label{eqn:clusterCoefficients}
	\braket{s_1,\dots,s_n}{Cl_n}=2^{-n/2} (-1)^p,
\end{equation}
where $p$ denotes the number of sites $i$ such that $s_i=s_{i+1}=1$.

This observation makes it simple to derive the tensors of the MPS
representation. We need a $D=2$-dimensional correlation system, which
-- loosely speaking -- will convey the information about the state
$s_i$ of the $i$-th site to site $i+1$. 
Define the matrices $A[0/1]$ by 
\begin{eqnarray}
	\label{eqn:clusterMatrices}
	\label{eqn:clusterMatrix0}
	\xminid{\ar[r]&\xnode{A[0]}\ar[r]&}&=&\ket+_r \bra0_l, 
	\\
	\label{eqn:clusterMatrix1}
	\xminid{\ar[r]&\xnode{A[1]}\ar[r]&}&=&\ket-_r \bra1_l.  
\end{eqnarray}
The intuition behind this choice is as follows. By the elementary
relations
\begin{equation}
	\braket+0=\braket+1=\braket-0=2^{-1/2},\qquad \braket-1=-2^{-1/2},
\end{equation}
the contraction in the middle of
\begin{equation}	
	\xminid{\ar[r]&\xnode{A[s_1]}\ar@{-}[r]&\xnode{A[s_2]}\ar[r]&} 
\end{equation}
will
yield a sign of "$-1$" exactly if $s_1=s_2=1$. Indeed, setting the
boundary vectors to $\ket L=\ket0, \ket R=\ket+$ one checks easily
that
\begin{equation}
	\bra R A[s_n]\dots A[s_1]\ket L = 2^{-n/2} (-1)^p,
\end{equation}
which is exactly the value required by Eq.
(\ref{eqn:clusterCoefficients}).

%Irrespective of the initial state $\ket\psi$ of the correlation system, the
%output of $A[s_i]$ is always proportional to $\ket{+}$ (in case of
%$s_i=0$) or $\ket-$ (for $s_i=1$):
%\begin{equation}\label{eqn:prepare}
%	\xminid{\xnode{\psi}\ar[r]&\xnode{A[s_i]}\ar[r]&}
%	\propto
%	\left\{
%		\begin{array}{ll}
%			\ket+ \quad \text{if } s_{i+1}=0\\
%			\ket- \quad \text{if } s_{i+1}=1.
%		\end{array}
%	\right.
%	%\quad \forall\, \ket\psi.
%\end{equation}
%Hence, to the right of the $i$-th site, the correlation system carries
%the information about $s_i$, as required.
%
%Likewise, 
%\begin{equation}\label{eqn:measure}
%	\xminid{\ar[r]&\xnode{A[s_{i+1}]}\ar[r]&\xnode{\psi}} \propto \bra{s_i},
%\end{equation}
%We see that the concatenation of two such tensors
%\begin{equation}
%	\xminid{\ar[r]&\xnode{A[s_i]}\ar[r]&\xnode{A[s_{i+1}]}\ar[r]&}
%\end{equation}	
%introduces a phase of $-1$ whenever $s_1=s_2=1$. This intuition can be
%made more precise by setting $\ket L=\ket0, \ket R=\ket+$ and checking
%that
%\begin{equation}
%	\bra R A[s_n]\dots A[s_1]\ket L = 2^{-n} (-1)^p,
%\end{equation}
%exactly the value required by Eq.\ (\ref{eqn:clusterCoefficients}).

Below, we will interpret the correlation system of a 1-D chain as a
single logical quantum system. For this interpretation to be viable,
we must check that the following basic operations can be performed
deterministically by local measurements: i) prepare the correlation
system in a known initial state, ii) transport that state along the
chain (possibly subject to known unitary transformations) and iii)
read out the final state.

To set the state of the correlation system to a definitive value, we
measure some site -- say the $i$-th -- in the $Z$-eigenbasis. 
Throughout this work, we will choose the notation 
$X$, $Y$, and $Z$ for the {\it Pauli operators}. Denote
the measurement outcome by $z\in\{0,1\}$.  In case of $z=0$, Eq.\
(\ref{eqn:clusterMatrix0}) tells us that the state of the correlation
system to the right of the $i$-th site will be $\ket+$ (up to an
unimportant phase). Likewise, a $z=1$ outcome prepares the correlation
system in $\ket-$, according to Eq.\ (\ref{eqn:clusterMatrix1}).  It
follows that we can use $Z$-measurements for preparation.  How to cope
with the intrinsic randomness of quantum measurements will concern us
later.

Secondly, consider the operators
\begin{eqnarray}
	\xymatrix@C=3mm{\ar[r]&\xnode{A[+]}\ar[r]&}
	&=&
	2^{-1/2}(
		\xymatrix@C=3mm{\ar[r]&\xnode{A[0]}\ar[r]&}
		+
		\xymatrix@C=3mm{\ar[r]&\xnode{A[1]}\ar[r]&}
	) \nonumber \\
	&\propto&\ket+\bra0+\ket-\bra1=H, 
	\label{eqn:clusterTransport+} \\
	\nonumber \\
	\xymatrix@C=3mm{\ar[r]&\xnode{A[-]}\ar[r]&}&\propto&H Z,
	\label{eqn:clusterTransport-} 
\end{eqnarray}
where $H$ is the Hadamard-gate.  We see immediately that measurements
in the $X$-eigenbasis give rise to a unitary evolution on the correlation
space. Similarly, one can show that one can generate arbitrary local 
unitaries by appropriate measurements in the $Y$-$Z$ plane.

Below, we will frequently be confronted with a situation like the one
presented in Eqs.\
(\ref{eqn:clusterTransport+},\ref{eqn:clusterTransport-}), where the
correlation system evolves in one of two possibilities, dependent
on the outcome of a measurement. It will be convenient to introduce a
compact notation that encompasses both cases in a single equation.
So Eqs.\ (\ref{eqn:clusterTransport+},\ref{eqn:clusterTransport-}) 
will be represented as 
\begin{equation}\label{eqn:boxedTransport}
	\xminid{\ar[r]&\xnode{A[X]}\ar[r]&}=H Z^x.
\end{equation}
Here $x=0$ corresponds to the outcome $|+\rangle $ in an 
$X$-measurement, whereas $x=1$ corresponds to the outcome $|-\rangle$.
In general, a physical observable given as an argument to a
tensor corresponds to a measurement in the observable's eigenbasis.
The measurement outcome is assigned to a suitable variable as in the
above example. 

Lastly, we must show how to physically read out the state of the
purely logical correlation system.  It turns out that measuring the
$i+1$-th physical system in the $Z$-eigenbasis corresponds to a
$Z$-measurement of the state of the correlation system just after site
$i$. Indeed, suppose we have measured the first $i$ systems and
obtained results corresponding to the local projection operator
$\ket{\phi_1}\otimes\dots\otimes\ket{\phi_i}$. Further assume that as
a result of these measurements the correlation system is in the 
state $\ket0$:
\begin{equation}
	\xymatrix@C=1mm@!{
		*+[F]{L}\ar@{-}[r]&*+[F]{A[\phi_1]}\ar@{-}[r]&\dots\ar@{-}[r]&*+[F]{A[\phi_i]}\ar[r]&
	}=\ket0.
\end{equation}
Using Eq.\ (\ref{eqn:clusterMatrix1}) we have that
%$\xminid{\ar[r]&\xnode{A[1]}}\propto \bra1$ and hence
\begin{eqnarray}
	&&
	\xymatrix@C=2mm@!{
		*+[F]{L}\ar@{-}[r]&*+[F]{A[\phi_1]}\ar@{-}[r]&\dots\ar@{-}&*+[F]{A[\phi_i]}\Par&
		*+[F]{A[1]}\ar[r]&
	} \\
	&\propto& \ket+ \braket 10=0.\nonumber
\end{eqnarray}
But then it follows from Eq.\ (\ref{eqn:transport}) that the
probability of obtaining the result $1$ for a $Z$-measurement on site
$i+1$ is equal to zero. In other words: if the \emph{correlation
system} is in the state $\ket0$ after the $i$-th site, then the $i+1$-th
\emph{physical site} must also be in the state $\ket0$. An analogous
argument for the $\ket1$-case completes the description of the
read-out scheme.

\subsection{2-D lattices}
\label{sec:2d}

The graphical notation greatly facilitates the passage to 2-D
lattices.  Here, the tensors $A[s]$ have four indices
$A[s]_{l,r,u,d}$, which will be contracted with the indices of the
left, right, upper and lower neighboring tensors respectively. After
choosing a set of boundary conditions $\ket L, \ket R, \ket U, \ket D
\in \cc^D$, the expansion coefficients of the state vector $\ket\Psi$
are computed as illustrated in the following example on a $2\times
2$-lattice:
\begin{eqnarray} \label{eqn:2d}
\braket{s_{1,1}, \dots, s_{2,2}}{\Psi}=
%	\begin{xy}
%	*!C\xybox{\xymatrix@C=5mm@R=3mm{
%					        &    \xnode{U}        & \xnode{U}     \\
%%	 \xnode{L}\ar[r]&\xnode{s_{1,1}}\ar[u]\ar[r]&\xnode{s_{2,1}}\ar[u]\ar[r]&\xnode{R} \\
%%	 \xnode{L}\ar[r]&\xnode{s_{1,2}}\ar[u]\ar[r]&\xnode{s_{2,2}}\ar[u]\ar[r]&\xnode{R} \\
%	 \xnode{L}\ar[r]&\xnode{A[s_{1,1}]}\ar[u]\ar[r]&\xnode{A[s_{2,1}]}\ar[u]\ar[r]&\xnode{R} \\
%	 \xnode{L}\ar[r]&\xnode{A[s_{1,2}]}\ar[u]\ar[r]&\xnode{A[s_{2,2}]}\ar[u]\ar[r]&\xnode{R} \\
%					        &\xnode{D}\ar[u]      &\xnode{D}\ar[u]     
%	}}
%	\end{xy}
	\begin{xy}
	*!C\xybox{\xymatrix@C=3mm@R=3mm{
					        &    \xnode{U}        & \xnode{U}     \\
	 \xnode{L}\Par&\xnode{A[s_{1,1}]}\Pau\Par&\xnode{A[s_{2,1}]}\Pau\Par&\xnode{R} \\
	 \xnode{L}\Par&\xnode{A[s_{1,2}]}\Pau\Par&\xnode{A[s_{2,2}]}\Pau\Par&\xnode{R} \\
					        &\xnode{D}\Pau      &\xnode{D}\Pau     
	}}
	\end{xy}.
\end{eqnarray}

In the 1-D case, we thought of the quantum information as
moving along a single correlation system from the left to the right.
For higher-dimensional lattices,
a greater deal of flexibility proves to be expedient. For example, 
sometimes it will be natural to interpret the tensor $A_{l,r,u,d}$ as specifying
the matrix elements of an operator $A$ mapping the left and the lower
correlation systems to the right and the upper ones:
\begin{equation}
 A_{l,r,u,d}=\bra r\otimes \bra u\,A\,\ket l\otimes\ket d,
 \quad
 A=
	\begin{xy}
	*!C\xybox{\xymatrix@C=3mm@R=3mm{
	 &&\\
	 \ar[r]&\xnode{A}\ar[u]\ar[r]& \\
	 &\ar[u]&\\
	}}
	\end{xy}.
\end{equation}
Often, on the other hand, the interpretation 
\begin{equation}
 A_{l,r,u,d}=\bra r\,A\,\ket l\otimes\ket u\otimes\ket d,
 \quad
 A=
	\begin{xy}
	*!C\xybox{\xymatrix@C=3mm@R=3mm{
	 &\ar[d]&\\
	 \ar[r]&\xnode{A}\ar[r]& \\
	 &\ar[u]&\\	
	}}
	\end{xy}
\end{equation}
or yet another one is to be preferred.

We have seen in Section \ref{1DTN} that the correlation system of a
one-dimensional matrix product state can naturally be interpreted as a
single quantum system subject to a time evolution induced by local
measurements. It would be desirable to carry this intuition over to
the 2-D case. Indeed, most of the examples to be discussed below are all
similar in relying on the same basic scenario: some horizontal lines
in the lattice are interpreted as effectively one-dimensional systems,
in which the logical qubits travel from the left to the right.
The vertical dimension is used to either couple the logical systems or
isolate them from each other (see Fig.\ \ref{fig:flow}). The reader
should recall that this setting is very similar to the original
cluster state based-techniques. Clearly, it would be interesting to devise
schemes not working in this way and the example presented in Section
\ref{sec:toric2} takes a first step in 
this direction.

\subsection{Example: the 2-D cluster state}
\label{sec:2dcluster}

Once again the cluster state serves as an example. One can work out
the tensor network representation of the 2-D cluster state vector
$\ket{Cl_{n\times n}}$ in the same way utilized for the 1-D case in
Section \ref{sec:1dcluster}. The resulting tensors are:
\begin{eqnarray}
	\label{eqn:2dcluster0}
	\begin{xy}
	*!C\xybox{\xymatrix@C=3mm@R=3mm{
	 &&\\
	 \ar[r]&\xnode{A[0]}\ar[u]\ar[r]& \\
	 &\ar[u]&
	}}
	\end{xy}
	=\ket+_r\ket+_u\,\bra0_l\bra0_d, \\
	\label{eqn:2dcluster1}
	\begin{xy}
	*!C\xybox{\xymatrix@C=3mm@R=3mm{
	 &&\\
	 \ar[r]&\xnode{A[1]}\ar[u]\ar[r]& \\
	 &\ar[u]&\
	}}
	\end{xy}
	=\ket-_r\ket-_u\,\bra1_l\bra1_d, \\
	%\xminid{&\xnode{L}\ar[r]&}=\ket L,\quad
	%\begin{xy}
	%*!C\xybox{\xymatrix@C=3mm@R=3mm{
	% \\
	% \xnode{D}\ar[u]
	%}} 
	%\end{xy}
	%= \ket D,\quad
	\ket L=\ket D =\ket+,
	\qquad
	\ket R=\ket U=\ket1.
\end{eqnarray}
An important property of Eqs.\ (\ref{eqn:2dcluster0},
\ref{eqn:2dcluster1}) is that the tensors $A[0/1]$ factor. One could
graphically represent this fact by writing
\begin{equation}\label{eqn:factor}
	\begin{xy}
	*!C\xybox{\xymatrix@C=3mm@R=3mm{
	 &&\\
	 \Par&\xnode{A[0]}\Pau\Par& \\
	 &\Pau&
	}}
	\end{xy}
	=
	\begin{xy}
	*!C\xybox{\xymatrix@C=3mm@R=3mm{
	 &&&&\\
	 &&\xnode{+}\Pau&\\
	 \Par&\xnode{0}&&\xnode{+}\Par&\\
	 &&\xnode{0}&\\
	 &&\Pau
	}}
	\end{xy},
\end{equation}
where 
\begin{equation}
	\xminid{\xnode{0}\ar[r]&}=\ket0,\,\,
	\xminid{\xnode{+}\ar[r]&}=\ket+. 
\end{equation}
In other words: the tensors
$A[0/1]$ effectively de-couple their respective indices. Based on this
fact, we will see momentarily how $Z$-measurements can be used to stop
information from flowing through the lattice. 

Indeed, suppose three vertically adjacent sites are measured, from top to
bottom, respectively in the $Z$, $X$ and $Z$-eigenbasis:
\begin{equation}
	\begin{xy}
	*!C\xybox{\xymatrix@C=3mm@R=3mm{
		&&&\\
	 	\ar[r]&\xnode{A[Z_u]}\ar[u]\ar[r]&\\
	 	\ar[r]&\xnode{A[X]}\ar@{-}[u]\ar[r]&\\
	 	\ar[r]&\xnode{A[Z_d]}\ar@{-}[u]\ar[r]&\\
	  &\ar[u]&
	}}
	\end{xy}.
\end{equation}
Denote the measurement results by $z_u, x, z_d\in\{0,1\}$.
As before, these numbers correspond to $z_u=0$ for 
$|0\rangle$ and $z_u=1$ for $|1\rangle$, as well as
$x=0$ for $|+\rangle$ and $x=1$ for $|- \rangle$.
In fact, we are mainly interested in the indices of the middle tensor,
as they will be the ones which carry the logical information. 
To this end Eq.\ (\ref{eqn:factor}) is of use, as it says that the 
upper and lower
tensors factor and hence it makes sense to dis-regard all of their
indices which do not influence the middle part. It hence suffices to
consider
\begin{equation}
	\begin{xy}
	*!C\xybox{\xymatrix@C=3mm@R=3mm{
	 	&\xnode{A[Z_u]}\ar@{-}[d]&\\
	 	\ar[r]&\xnode{A[X]}\ar[r]&\\
	 	&\xnode{A[Z_d]}\ar@{-}[u]&\\
	}}
	\end{xy}.
\end{equation}
As a first step, we calculate
\begin{eqnarray*}
	\begin{xy}
	*!C\xybox{\xymatrix@C=3mm@R=3mm{
	 	&\xnode{0}\ar@{-}[d]&\\
	 	\ar[r]&\xnode{A[0]}\ar[r]&\\
	 	&\xnode{+}\ar@{-}[u]&\\
	}}
	\end{xy}
	&=&	
	\begin{xy}
	*!C\xybox{\xymatrix@C=3mm@R=3mm{
	 &&\xnode{0}\Pad&\\
	 &&\xnode{+}\\
	 \ar[r]&\xnode{0}&&\xnode{+}\ar[r]&\\
	 &&\xnode{0}\Pad&\\
	 &&\xnode{+}
	}}
	\end{xy}
	=2^{-1} \ket+\bra0,
\end{eqnarray*}
having used Eq.\ (\ref{eqn:factor}) and the basic fact
\begin{equation}
	\xminid{\xnode{+}\Par[r]&\xnode{0}}=\braket0+=2^{-1/2}. 
\end{equation}
A similar
calculation where $A[0]$ is substituted by $A[1]$ yields 
$2^{-1}
\ket-\bra1$. Hence, for $A[+]\propto A[0]+A[1]$, we have
\begin{eqnarray}
	\begin{xy}
	*!C\xybox{\xymatrix@C=3mm@R=3mm{
	 	&\xnode{0}\ar@{-}[d]&\\
	 	\ar[r]&\xnode{A[+]}\ar[r]&\\
	 	&\xnode{+}\ar@{-}[u]&\\
	}}
	\end{xy}
	&\propto&	
	\ket+\bra0+\ket-\bra1=H.
\end{eqnarray}
Similarly,
\begin{eqnarray}
	\begin{xy}
	*!C\xybox{\xymatrix@C=3mm@R=3mm{
	 	&\xnode{0}\Pad&\\
	 	\ar[r]&\xnode{A[-]}\ar[r]&\\
	 	&\xnode{+}\Pau&\\
	}}
	\end{xy}
	&\propto&	
	H Z.
\end{eqnarray}
After these preparations it is simple to conclude that
\begin{equation} \label{eqn:2dtransport}
	\begin{xy}
	*!C\xybox{\xymatrix@C=3mm@R=3mm{
	 	&\xnode{A[Z_u]}\ar@{-}[d]&\\
	 	\ar[r]&\xnode{A[X]}\ar[r]&\\
	 	&\xnode{A[Z_d]}\ar@{-}[u]&\\
	}}
	\end{xy}
	\propto H Z^{z_u+x+z_d}.
\end{equation}
This finding tells us how to transport quantum information along
horizontal lines through the lattice. Namely by measuring the line in
the $X$-eigenbasis to cause the information to flow from the left to
the right and measuring vertically adjacent sites in the
$Z$-eigenbasis to shield the information from the rest of the lattice.

Eq.\ (\ref{eqn:2dtransport}) should be compared with Eqs.\
(\ref{eqn:clusterTransport+},\ref{eqn:clusterTransport-}). So up to
possible corrections of the form $Z^{z_u+z_l}$, the procedure outlined
above enables us to effectively prepare a 1-D cluster state within the
2-D lattice. 

\section{Novel resource states}
\label{sec:novelty}

Up to this point, we have reformulated the computational model of the
one-way computer in the language of computational tensor networks.
This picture of one-way computation is educational in its own right.
However, to convincingly argue that the framework is rich enough to
allow for quite different models, we have to explicitly construct
novel schemes. It is the purpose of this section to discuss a number
of examples of new resources. As before, important features will be
highlighted as ``observations''.

\subsection{AKLT-type states}
\label{sec:aklt}

\subsubsection{1-D structures}
\label{sec:1daklt}

Our first example is inspired by the {\it AKLT state} \cite{aklt},
which is well-known in the context of condensed matter physics. The
{\it AKLT model} is a 1-D, spin-1, nearest neighbor, frustration free,
gapped Hamiltonian. Its unique ground state is a matrix product state
with $D=2$ and indeed, the AKLT model motivated the first studies of
such states \cite{aklt,FCS}. The defining matrices of the MPS description
are:
\begin{eqnarray}
	\label{eqn:akltOrg}
	\xminid{\ar[r]&\xnode{A[0]}\ar[r]&} &=& Z,\\
	\xminid{\ar[r]&\xnode{A[1]}\ar[r]&} &=& 
		2^{-1/2}
	|0\rangle_r\langle 1|_l,\\
	\xminid{\ar[r]&\xnode{A[2]}\ar[r]&} &=& 
		2^{-1/2} 
		|1\rangle_r\langle 0|_l
\end{eqnarray}
We will choose the boundary conditions to be $\ket L=\ket R=\ket0$. 
As a matter of fact, we will not work directly with the AKLT state,
but with a small variation, for which it turns out to be more
straight-forward to construct a scheme for MBQC. In this modification,
the matrix $A[0]$ is given by the Hadamard gate, instead of the Pauli
$Z$ operator:
\begin{equation}\label{eqn:akltH}
	\xminid{\ar[r]&\xnode{A[0]}\ar[r]&} = H.
\end{equation}
This state shares all the defining properties of the original: it is
the unique ground-state of a spin-1 nearest neighbor frustration free
gapped Hamiltonian (see Appendix \ref{sec:akltAppendix}). Against the
background of our program, the obvious question to ask is whether
these matrices can be used to implement any evolution on the
correlation space. 

To show that this is indeed the case, let us first analyze a
measurement in the $\{\ket0, \ket+, \ket-\}$-basis, where
$\ket\pm:=2^{-1/2}(\ket1\pm\ket2)$. In a mild abuse of notation, we will
hence write $\ket\pm$ for state vectors in the subspace spanned
by $\{ \ket1,\ket2\}$ instead of $\{ \ket0 ,\ket1\}$. From Eqs.\
(\ref{eqn:akltOrg}-\ref{eqn:akltH}) one finds that depending on the
measurement outcome, the operation realized on the correlation space
will be one of $H, X$ or $ZX=i\,Y$.  At this point, we have to turn to
an important issue: how to compensate for the randomness of quantum
measurement outcomes.  

\subsubsection{Compensating the randomness}
\label{sec:randomness}

Assume for
now that we intended to just transport the information faithfully from
left to right. In this case, we consider the operator 
\begin{equation}
	B_1:=H, X,\>\text{or}\>ZX
\end{equation}	
as an unwanted \emph{by-product} of the scheme. The one-way computer
based on cluster states has the remarkable property that the
by-products can be dealt with by adjusting the measurement-bases
depending on the previous outcomes, without changing the general
``layout'' (in the sense of Fig.\ \ref{fig:flow}) of the computation
\cite{OnewayCompModel}.  For more general models, as the ones
considered in this work, such a simple solution seems not available.
Fortunately, we can employ a ``trial-until-success'' strategy, which
proves remarkably general.

The key points to notice are that i) the three possible outcomes $H,
X$ and $Z$ generate a finite group $\mathcal{B}$ and ii) the
probability for each outcome is equal to $1/3$, independent of the
state of the correlation system. We will refer to $\mathcal{B}$ as the
model's \emph{by-product} group. Now suppose we measure $m$ adjacent
sites in the $\{\ket0,\ket+,\ket-\}$-basis. The resulting overall
by-product operator $B=B_m B_{m-1}\dots B_1$ will be a product of $m$
generators $H, X, ZX$.  So by repeatedly transporting the state of the
correlation system to the right, the by-products are subject to a
random walk on $\mathcal{B}$. Because $\mathcal{B}$ is finite, every
element will occur after a finite expected number of steps (as one can
easily prove).

The group structure opens up a way of dealing with the randomness.
Indeed, assume that initially the state vector of the correlation
system is given by $B \ket\psi$, for some unwanted $B\in \mathcal{B}$.
Transferring the state along the chain will introduce the additional
by-product operator $B^{-1}$ after some finite expected number of
steps, leaving us with 
\begin{equation} 
	B^{-1} B \ket\psi=\ket\psi,
\end{equation}
as desired. The technique outlined here proves to be extremely general
and we will encounter it in further examples presented below.

\begin{observation}[Compensating 
randomness]\label{obs:randomness} 
Possible sets of by-product operators are not limited to the Pauli
group.  A way of compensating randomness for other finite by-product
operator groups is to adopt a ``trial-until-success strategy'', which
gives rise to a random length of the computation.  This length is in
each case shown to be bounded on average by a constant in the system
size.
\end{observation}

\subsubsection{All single-qubit gates}

By the preceding paragraphs, we can implement any element of
$\mathcal{B}$ on the correlation space. 
We next address the problem of realizing a phase gate
$S(\phi):=\operatorname{diag}(1,e^{i\phi})$ for some
$\phi\in\mathbb{R}$.
To this end, consider a measurement on the
$\{\ket0,2^{-1/2}(\ket1\pm e^{i\phi}\ket2)\}$-basis. There are three
cases
\begin{itemize}
	\item
	The outcome corresponds to $\ket1+e^{i\phi}\ket2$. In this case, we
	get $S(\phi)$ on the correlation space and are hence done.

	\item
	The outcome corresponds to $\ket1-e^{i\phi}\ket2$. We get $Z S(\phi)$, 
	which is the desired operation, up to an element of the
	by-product group, which we can rid ourselves of as described above.

	\item
	Lastly, in case of $\ket0$, we implement $H$ on the correlation
	space. As $H\in \mathcal{B}$, we can ``undo'' it and then re-try to
	implement the phase gate.
\end{itemize}

Hence, we can implement any element of ${\cal B}$ as well as $S(\phi)$
on the correlation space. This implies that $H S(\phi) H$ is also
realizable and therefore any single-qubit unitary, as $SU(2)$ is
generated by operations of the form $S(\phi)$ and $HS(\phi)H$.

The state of the correlation system can be prepared by measuring in
the computational basis. In case one obtains a result of ``$1$'' or
``$2$'', the state of the correlation system will be $\ket0$ or
$\ket1$ respectively, irrespective of its previous state. A
``$0$''-outcome will not leave the correlation system in a definite
state. However, after a finite expected number of steps, a measurement
will give a non-``0''-result. Lastly, a read-out scheme can be
realized similarly (c.f.\ Section \ref{sec:1dcluster}).

\begin{observation}[Ground states]
	Ground states of one-dimensional gapped
	nearest-neighbor Hamiltonians may 
	serve as resources for
	transport and arbitrary rotations.
\end{observation}

\subsubsection{2-D structures}
\label{sec:2daklt}
\label{sec:aklt2D}

\begin{figure}
	\includegraphics[width=6.5cm]{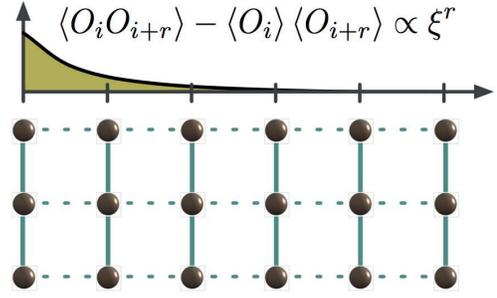}
	\caption{\label{fig:aklt}
		A universal resource deriving from the AKLT-model.}
\end{figure}

Several horizontal 1-D AKLT-type states can be coupled to become a
universal 2-D resource. The coupling can be facilitated by performing
a controlled-Z operation, embedded into the three-dimensional
spin-1 space, between vertically adjacent nearest neighbors. More
specifically, we will use the operation $\exp\{i \pi
\ket2\bra2\otimes\ket2\bra2\}$, which introduces a $\pi$-phase
between two systems exactly if both are in the state $\ket2$. The
tensor network representation of this resource is given by
\begin{eqnarray}
	\begin{xy}
	*!C\xybox{\xymatrix@C=3mm@R=3mm{
	 &&\\
	 \ar[r]&\xnode{A[0]}\ar[u]\ar[r]& \\
	 &\ar[u]&
	}}
	\end{xy}
	&=&H_{l\to r}\otimes \ket+_u\bra0_d, \\
	\begin{xy}
	*!C\xybox{\xymatrix@C=3mm@R=3mm{
	 &&\\
	 \ar[r]&\xnode{A[1]}\ar[u]\ar[r]& \\
	 &\ar[u]&\
	}}
	\end{xy}
	&=&2^{-1/2} \ket0_r\bra1_l\otimes \ket+_u\bra0_d, \\
	\begin{xy}
	*!C\xybox{\xymatrix@C=3mm@R=3mm{
	 &&\\
	 \ar[r]&\xnode{A[2]}\ar[u]\ar[r]& \\
	 &\ar[u]&\
	}}
	\end{xy}
	&=&2^{-1/2} \ket1_r\bra0_l\otimes  \ket-_u\bra1_d,
\end{eqnarray}
as one can check in analogy to Sec.\ \ref{sec:2dcluster}. Here,
\begin{equation}
	H_{l\to r}:=\ket+_r\bra0_l+\ket-_r\bra1_l.
\end{equation}

To verify that the resulting 2-D state constitutes a universal
resource, we need to check that a) one can isolate the correlation
system of a horizontal line from the rest of the lattice, so that it
may be interpreted as a logical qubit and b) one can couple these
logical qubits to perform an entangling gate. 

The first step works in complete analogy to Section
\ref{sec:2dcluster}, see Fig.\ \ref{fig:aklt}. 
Indeed, one simply confirms that
\begin{equation}
	\begin{xy}
	*!C\xybox{\xymatrix@C=3mm@R=3mm{
	 	&\xnode{A[Z_u]}\ar@{-}[d]&\\
	 	\ar[r]&\xnode{A[s]}\ar[r]&\\
	 	&\xnode{A[Z_l]}\ar@{-}[u]&\\
	}}
	\end{xy}
	= \pm \xminid{\ar[r]&\xnode{A[s]}\ar[r]&},
\end{equation}
where $s\in\{0,1,2\}$ and $Z_{u/l}$ denotes a measurement in the
$\{\ket0,\ket1,\ket2\}$-basis.  So measuring the vertically adjacent
nodes in the computational basis gives us back the 1-D state, up to a
possible sign. 
%The sign may introduce additional $Z$ errors on the
%correlation system, which can easily be dealt with, as $Z\in{\cal B}$.

A controlled-$Z$ gate can be realized in five steps:
\begin{eqnarray}
	\begin{xy}
	*!C\xybox{\xymatrix@C=5mm@R=3mm{
	&-2&-1&0&1&2\\
	\ar[r]&\xmnode{X}\ar@{-}[r]& \xmnode{X}\ar@{-}[r]&\xmnode{X}\ar@{-}[r] &\xmnode{X}\ar@{-}[r]&\xmnode{X}\ar[r]& \\
	\ar[r]&\xmnode{Z}\ar@{-}[r]\ar@{-}[u]\ar@{-}[d]&\xmnode{Z}\ar@{-}[u]\ar@{-}[d]\ar@{-}[r]&\xmnode{Y}\ar@{-}[u]\ar@{-}[d]\ar@{-}[r]&
	  \xmnode{Z}\ar@{-}[u]\ar@{-}[d]\ar@{-}[r]&\xmnode{Z}\ar@{-}[u]\ar@{-}[d]\ar[r]&\\
	\ar[r]&\xmnode{X}\ar@{-}[r]& \xmnode{X}\ar@{-}[r]&\xmnode{X}\ar@{-}[r] &\xmnode{X}\ar@{-}[r]&\xmnode{X}\ar[r]& 
	}}
	\end{xy}.\nonumber\\
\end{eqnarray}
The Pauli matrices $X,Y,Z$ are understood as being embedded into the
$\{\ket1,\ket2\}$-subspace. So, e.g., $X$ denotes a measurement in the
$\{\ket0,2^{-1/2}(\ket1\pm\ket2)\}$-basis. When operating the gate, 
we first measure all sites of the upper and lower lines in the
$X$-eigenbasis. In case the result for the sites at position
``0'' (refer to labeling above) is different from $\ket+$, the gate
failed. In that case all sites on the middle line are measured in the
computational basis and we restart the procedure five steps to the
right\footnote{
	We have chosen this approach in order to avoid an awkward discussion
	of how to handle phases introduced by ``wrong'' measurement
	outcomes. We are providing proofs of principle for universality
	here and will accept a (possibly daunting) linear overhead in the
	expected number of steps, if this simplifies the discussion.
	Substantial improvements to these schemes are, of course, possible.
}. 
Otherwise, the systems labeled by a $Z$ are measured. We accept the
outcome only if we obtained $\ket1$ on sites $\pm 2$ and $\ket0$ on
sites $\pm 1$ -- should a different result occur, the gate is once
again considered a failure and we proceed as above.  Lastly, the $Y$
measurement on the central site is performed. In case of a result
corresponding to $\ket0$, it is easy to see that no interaction
between the upper and the lower part takes place, so this is the last
possibility for the gate to fail. Let us assume now that the desired
measurement outcomes were realized. At site $-2$ on the middle line,
we obtained 
\begin{equation}
	\xminid{\xnode{A[1]}\ar[r]&},
\end{equation}
which prepares the
correlation system of the middle line in $\ket0$. At site $-1$, in
turn, a Hadamard gate has been realized, which causes the output of
site $-1$ to be $H\ket0=\ket+$. The situation is similar on the
r.h.s., so that the above network at site $0$ can be re-written as
\begin{equation}\label{eqn:akltEntangling}
	\begin{xy}
	*!C\xybox{\xymatrix@C=5mm@R=3mm{
	\ar[r]						&\xmnode{+}\ar[r]& \\
	\xnode{+}\Par 		&\xmnode{Y}\Par\ar@{-}[u]& \xnode{+}\\
	\ar[r]						&\xmnode{+}\ar[r]\ar@{-}[u]& \\
	}}
	\end{xy}.
\end{equation}
We will now analyze the tensor network in Eq.\
(\ref{eqn:akltEntangling})
step by step. For proving its functionality, there is no loss of
generality in restricting attention to the situation where the
correlation system of the lower line is initially in state $\ket c$,
for $c\in\{0,1\}$. We compute for the lower part of the tensor network
\begin{equation}
	\begin{xy}
	*!C\xybox{\xymatrix@C=5mm@R=3mm{
	&&\\
	\xnode{\ket c}\ar@{-}[r]&\xmnode{+}\ar[r]\ar[u]& 
	}}
	\end{xy}
	=X\ket c_r Z^c \ket+_u.
\end{equation}
Further, plugging the output $Z^c\ket+$ of the lower stage into the
middle part, we find
\begin{equation}
	\begin{xy}
	*!C\xybox{\xymatrix@C=5mm@R=3mm{
	&&\\
	\xnode{+}\Par 	
	&\xmnode{Y}\Par\ar[u]& \xnode{+}\\
	&\xnode{Z^c\ket+}\ar@{-}[u] }}
	\end{xy}
	\propto Z^{c+y} (\ket0+i\ket1),
\end{equation}
where $y\in 0,1$ reflects the outcome of the $Y$-measurement on the
central site: $y=0$ in case of $\ket1+i\ket2$ and $y=1$ for
$\ket1-i\ket2$.
Lastly, 
\begin{equation}
	\begin{xy}
	*!C\xybox{\xymatrix@C=5mm@R=3mm{
	\ar[r]						&\xmnode{+}\ar[r]& \\
							&\xnode{Z^{c+y} (\ket0+i\ket1)}\ar@{-}[u]
	}}
	\end{xy}
	\propto S Z^{c+y} X.
\end{equation}
In summary, the evolution afforded on the upper line is $H S Z^{y+c}$,
equivalent to $Z^c$ up to by-products.  This completes the proof of
universality.

For completeness, note that we never need the by-products to vanish
for all logical qubits of the full computation 
simultaneously. Hence the expected number of
steps for the realization of one- or two-qubit gates is a 
constant in the number of total logical qubits.

\subsection{Toric code states}
\label{sec:toric}

% ttag

\newcommand{\xar}[1]{\ar@{-}[#1]}
\newcommand{\rem}[1]{\textbf{[#1]}}
\newcommand{\xmnodeA}[1]{*+[F]{A[#1]}}
\newcommand{\xmnodeB}[1]{*+[F]{B[#1]}}

In the following, we present two MBQC resource states which are motivated
by Kitaev's toric code states~\cite{kitaev:toriccodes}. This contrasts
with a result in Ref.\ \cite{bravyi:toric-mbc} that
MBQC on the planar toric code state itself can be
simulated efficiently classically.
Different from the other schemes presented, the natural gate in these
schemes is a two-qubit interaction, whereas local operations have to be
implemented indirectly. Also, individual qubits are decoupled not by
erasing sites but by switching off the coupling between them.

Toric code states are states with non-trivial topological properties 
and have been introduced in the context of quantum error correction.
They have a particularly simple representation in
terms of PEPS~\cite{frank:PEPS} or CTNs \cite{shortOne} 
on two centered square lattices, 
\vspace*{-1.8cm}
\begin{equation}
\label{eqn:nys:toriccode}
	\vspace*{-1.8cm}
	\begin{xy}
	*!C\xybox{\xymatrix@C=3mm@R=3mm@ru{
	 & &\xar{d} &\xar{d} & & & \\
	 &\xar{r}\xar{d} & \xnode{K_V}\xar{r}\xar{d} & 
	    \xnode{K_H}\xar{r}\xar{d} & \xar{d} &   & \\
	 \xar{r} & \xnode{K_V}\xar{r}\xar{d} & \xnode{K_H}\xar{r}\xar{d} &
	    \xnode{K_V}\xar{r}\xar{d} & \xnode{K_H}\xar{r}\xar{d} &
	    \xar{d} & \\
	 &\xar{r} & \xnode{K_V}\xar{r}\xar{d} & \xnode{K_H}\xar{r}\xar{d} &
	    \xnode{K_V}\xar{r}\xar{d} & \xnode{K_H}\xar{r}\xar{d} & \\
	 & & \xar{r} & \xnode{K_V}\xar{r}\xar{d} & 
	    \xnode{K_H}\xar{r}\xar{d} & & \\
	 & & & & & & 
	}}
	\end{xy}
\end{equation}
where 
\begin{eqnarray}
	\begin{xy}
	*!C\xybox{\xymatrix@C=3mm@R=3mm{
	\xar{rd} & & \\
	& \xnode{K_H[s]}\xar{ru}\xar{rd} & \\
	\xar{ru} & & 
	}}\end{xy}&=&
	\begin{xy}
	*!C\xybox{\xymatrix@C=3mm@R=3mm{
	    \xar{rd}& & \\
	    & \xnode{Z^s} \xar{ru} & \\
	    & \xnode {Z^s} \xar{rd} & \\
	    \xar{ru} & & 
	}}\end{xy}
\end{eqnarray}	
and
\begin{eqnarray}
	\begin{xy}
	*!C\xybox{\xymatrix@C=3mm@R=3mm{
	\xar{rd} & & \\
	& \xnode{K_V[s]}\xar{ru}\xar{rd} & \\
	\xar{ru} & & 
	}}\end{xy}&=&
	\begin{xy}
	*!C\xybox{\xymatrix@C=3mm@R=3mm{
	    \xar{rd}& & & \\
	    & \xnode{Z^s} & \xnode{Z^s}\xar{rd}\xar{ru} & \\
	    \xar{ru} & & &
	}}\end{xy}\ ,
\end{eqnarray}	
i.e., $K_H$ and $K_V$ are identical up to a rotation by $90$ degrees.

\begin{figure*}
\includegraphics[width=0.95\textwidth]{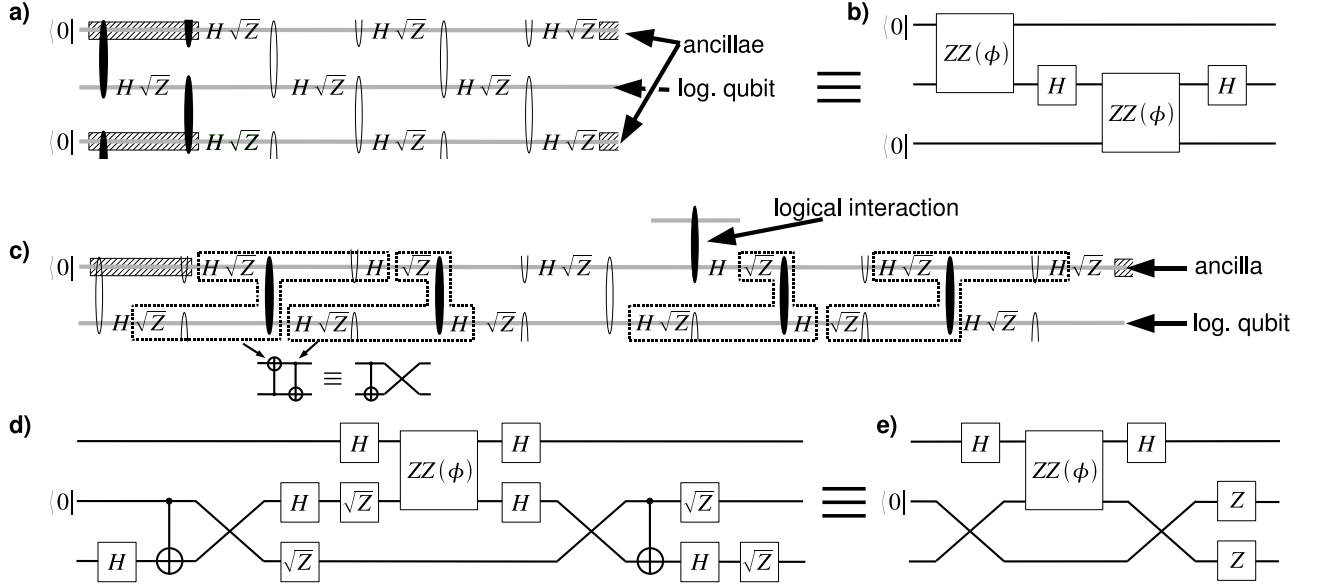}
\caption{Implementation of single-qubit and two-qubit operations in the
first toric code model. \textbf{a)} The measurement pattern for
single-qubit operations and \textbf{b)} the corresponding circuit.
\textbf{c)} Pattern for a two-qubit gate between logical qubits,
\textbf{d)} the corresponding circuit and \textbf{e)} the circuit after
some simplifications.}
\label{fig:toric1}
\end{figure*}

Let us first see how $K_H$ acts on two qubits in correlation space coming
from the left. The most basic operation is a measurement in the
computational basis, which simply transports both qubits to the right (up
to a correlated $Z$ by-product operator). Generalizing this to 
measurements in the $Y$-$Z$ plane, we find that
\begin{eqnarray}
	\begin{xy}
	*!C\xybox{\xymatrix@C=3mm@R=3mm{
	\ar[rd] & & \\
	& \xnode{K_H[\phi]}\ar[ru]\ar[rd] & \\
	\ar[ru] & & 
	}}\end{xy}&=&
	\begin{xy}
	*!C\xybox{\xymatrix@C=3mm@R=3mm{
	\ar[rd] & & \\
	& \xnode{ZZ(\phi)}\ar[ru]\ar[rd] & \\
	\ar[ru] & & 
	}}\end{xy}
\end{eqnarray}	
where $\phi$ is the angle with the $Z$ axis, and
\begin{equation}
ZZ(\phi)=\left(\begin{array}{cccc}1\\&e^{i\phi}\\&&e^{i\phi}\\&&&1
    \end{array}\right)\ .
\label{eq:nys:zz}
\end{equation}
(Note that this gate is locally equivalent to the 
\textsc{cnot} gate for $\phi=\pm\pi/2$.)

Thus, the tensors in Kitaev's toric code
state have a \emph{two}-qubit operation as their natural gate in
correlation space, rather than a
\emph{single}-qubit gate.  In MBQC schemes which base on these
projectors, two-qubit gates are easy to realize, whereas in order to get
one-qubit gates, tricks have to be used. In the first 
example, we obtain
single-qubit operations by introducing ancillae: a $ZZ$ controlled phase
between a logical qubit and an ancilla in a computational basis
state yields a local $Z$ rotation on the logical qubit.
In the second example, we use a
different approach: we encode each logical qubit in
\emph{two} qubits in correlation space. Using this nonlocal encoding, we
obtain an easy implementation of both one- and two-qubit operations;
furthermore, the scheme allows for an arbitrary parallelization of the
two-qubit interactions.

\begin{observation}[Logical qubits in several correlation
systems]\label{obs:kitaev}
	There is no need to have a one-one correspondance between logical
	qubits and a single correlation system.
\end{observation}

\subsubsection{Toric codes: first scheme}
\label{sec:toric1}

Our first scheme consists of the modified tensor
\begin{eqnarray}
	\begin{xy}
	*!C\xybox{\xymatrix@C=3mm@R=3mm{
	\ar[rd] & & \\
	& \xnode{\tilde K_H[s]}\ar[ru]\ar[rd] & \\
	\ar[ru] & & 
	}}\end{xy}&=&
	\begin{xy}
	*!C\xybox{\xymatrix@C=3mm@R=3mm{
	\ar[rd] & & \\
	& \xnode{K_H[s]}\ar[ru]\ar@{-}[rd] & &\\
	\ar[ru] & & \xnode{\sqrt{Z}H}\ar[rd]&\\
	&&&&
	}}\end{xy}
	\label{eq:nys:mod1}\\
%%%%%%%
&=&
	\begin{xy}
	*!C\xybox{\xymatrix@C=3mm@R=3mm{
	    \ar[rd]& & \\
	    & \xnode{\hspace*{1.2em}Z^s\hspace*{1.2em}} \ar[ru] & \\
	    & \xnode {\sqrt{Z}HZ^s} \ar[rd] & \\
	    \ar[ru] & & 
	}}\end{xy}\nonumber
\end{eqnarray}
[with $\sqrt{Z}=\mathrm{diag}(1,i)$], arranged as in 
(\ref{eqn:nys:toriccode}) where \emph{both} 
$K_H$ and $K_V$ are replaced by
$\tilde K_H$.  The extra $H$ serves the same purpose as in other
schemes: it allows to leave the subspace of diagonal operations and
thus to implement $X$ rotations. The need for the $\sqrt{Z}$ will become
clear later; it is connected to the fact that 
\begin{equation}
	\label{eq:nys:cnot}
	\mbox{\sc cnot}=(\id\otimes H)\;(\sqrt Z\otimes \sqrt{Z})\;
	ZZ(-\pi/2)\;(\id\otimes H)\ .
\end{equation}

In the following, we show how this state can be used for MBQC. The
qubits run from left to right in correlation space in zig-zag lines in
Eq.~(\ref{eqn:nys:toriccode}); for the illustration in 
Fig.~\ref{fig:toric1}, we have straightened these lines, and marked the
measurement-induced $ZZ$ interactions coming from 
the $K_H[s]$ in (\ref{eq:nys:mod1})
by ellipses. (The
difference between filled and non-filled ellipses will be explained
later.) The $\sqrt{Z}H$ operations of (\ref{eq:nys:mod1}) do not depend on
the measurement and are thus hard-wired; note that the
order is reversed as we are considering $H$ and $\sqrt Z$ as two
independent operations in the circuit.

Let us first impose that all qubits are
initialized to $\ket{0}$; this corresponds to a left boundary condition
$\ket{0}$ in correlation space. We will discuss later how to initialize the
scheme. Every second qubit is an ancilla which will be used
to implement one-qubit operations.  We first discuss the case of
no Pauli errors, and show later how those can be dealt with.

The implementation of single-qubit operations is illustrated in
Fig.~\ref{fig:toric1}a.  There, each ellipse denotes a possible $ZZ$
interaction. In particular, empty ellipses denote interactions which are
switched off (i.e.\ measured in the $Z$ basis), while filled ellipses
denote sites where one can measure in the $Y$-$Z$ plane to
implement a $ZZ$ gate.
If all interactions are switched off, all qubits are transported to the
right, subject to the transformation $\sqrt{Z}H$. As
$(\sqrt{Z}H)^3=\id$, the ancillae are in the computational basis in
every third step: These regions
are hashed in Fig.~\ref{fig:toric1}a. In these regions, a $ZZ(\phi)$
between ancilla and logical qubit (corresponding to the filled
ellipses in the figure) results in a single-qubit $Z$ rotation on the
latter. Thus, in each block of length three as the one shown in 
Fig.~\ref{fig:toric1}a, the transformation 
\begin{equation}
\sqrt{Z}H\sqrt{Z}HS(\psi)\sqrt{Z}HS(\phi)=
HS(\psi)HS(\phi)
\end{equation}
is implemented [where $S(\phi)=\mathrm{diag}(1,e^{i\phi})$], which allows
for arbitrary one-qubit operations.  In Fig.~\ref{fig:toric1}b, the
corresponding circuit is shown, which has been simplified using
$H\sqrt{Z}H\sqrt{Z}=\sqrt{X}\sqrt{Z}=(\sqrt{Z})^{-1}H$, and that diagonal
matrices commute.

Although the scheme has a natural two-qubit interaction, implementing an
interaction between two adjacent \emph{logical} qubits is complicated by
the ancilla which is located inbetween. In order to obtain a coupling, we
first swap the logical qubit with the ancilla, then couple it to the now
adjacent logical neighbor, and finally swap it back.  This is implemented
by the measurement pattern shown in Fig.~\ref{fig:toric1}c.  Again, empty
ellipses correspond to switched off interactions, while the filled
ellipses all implement $ZZ(-\pi/2)$ gates, each of which together with two
$\sqrt{Z}$ and two Hadamards as grouped in the figure gives a
\textsc{cnot} gate, cf.\ Eq.~(\ref{eq:nys:cnot}).  This measurement
pattern corresponds to the circuit shown in Fig.~\ref{fig:toric1}d,
where we have replaced
each pair of \textsc{cnot}s by a \textsc{cnot} and a \textsc{swap}. By
merging each \textsc{cnot} with the two adjacent Hadamards, we effectively
obtain 
\begin{equation}
	CZ = |0,0\rangle\langle 0,0 | + |0,1\rangle\langle 0,1|+
	|1,0\rangle\langle 1,0|  - |1,1\rangle\langle 1,1|
\end{equation}
gates. We thus remain with only
diagonal gates on the two lower qubits (except for the \textsc{swap}),
i.e.\ the gates all commute and the circuit can thus be simplified to the
one shown on in Fig.~\ref{fig:toric1}e, proving that the
sequence effectively implements a two-qubit interaction between the
logical qubits.  Note that the length of the complete sequence is
compatible with the three-periodicity of the basis of the ancillae.

Pauli errors in this scheme can be dealt with as usual: $H$ and
$\sqrt{Z}$ are both in the Clifford group, i.e., Paulis can be commuted
through, and $ZZ$ commutes with $Z$ errors, while $(\id\otimes
X)ZZ(\phi)=ZZ(-\phi)(\id\otimes X)$.

Finally, we show how to read out the logical qubits. It holds that
\renewcommand{\xar}[1]{\ar[#1]}
\begin{eqnarray}
	\begin{xy}
	*!C\xybox{\xymatrix@C=3mm@R=3mm{
	\xar{rd} & & \\
	& \xnode{H[+]}\xar{ru}\xar{rd} & \\
	\xar{ru} & & 
	}}\end{xy}&=&
    \left|\begin{array}{c}0\\0\end{array}\right\rangle
    \left\langle\begin{array}{c}0\\0\end{array}\right|+
    \left|\begin{array}{c}1\\1\end{array}\right\rangle
    \left\langle\begin{array}{c}1\\1\end{array}\right|\ ,
\\
	\begin{xy}
	*!C\xybox{\xymatrix@C=3mm@R=3mm{
	\xar{rd} & & \\
	& \xnode{H[-]}\xar{ru}\xar{rd} & \\
	\xar{ru} & & 
	}}\end{xy}&=&
    \left|\begin{array}{c}0\\1\end{array}\right\rangle
    \left\langle\begin{array}{c}0\\1\end{array}\right|+
    \left|\begin{array}{c}1\\0\end{array}\right\rangle
    \left\langle\begin{array}{c}1\\0\end{array}\right|\ ,
\end{eqnarray}	
i.e., a measurement in the $X$ basis returns the parity of the
ancilla and the logical qubit. If this is done when the ancilla is in
a computational basis state, one effectively measures the logical qubit in
the computational basis.  Note that both the ancilla and the logical qubit
are in a well-defined state afterwards and can thus be reused.

\begin{figure*}
\includegraphics[width=0.92\textwidth]{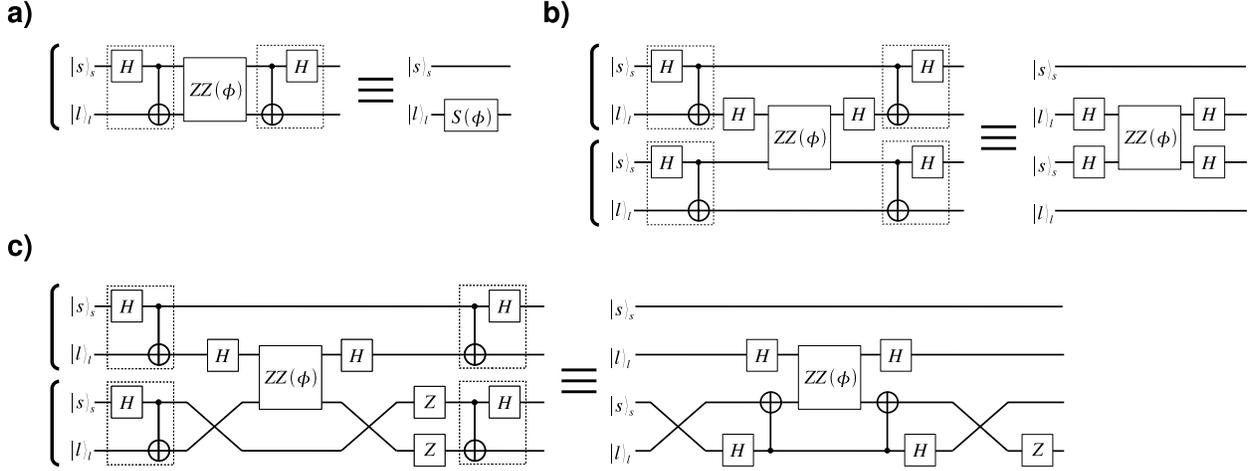}
\caption{Interpretation of the first toric code scheme in terms of parity
encoded qubits. The boxed parts of the circuit decode and encode the
system. \textbf{a)} $Z$ rotations result in $Z$ rotations in the encoded
system. \textbf{b)} $X$ rotations result in $X$ rotations in the encoded
system, plus $Z$ corrections before and after the rotations in case the
$s$ qubit below is $\ket{-}_s$ rather than $\ket{+}_s$. \textbf{c)}
Similarly, the coupling circuit Fig.~\ref{fig:toric1}d results in a
coupling of the encoded logical qubits, up to the same $Z$ correction on
the first logical qubit which depends on the $s$ qubit below in exactly
the same way. Thus, the $Z$ corrections on each qubit cancel out except
for the first and the last, which have no effect due to the initialization
and measurement in the computational basis.}
\label{fig:toric2}
\end{figure*}

Let us now turn towards the initialization procedure.  In contrast to 
the previous
MBQC schemes, the read-out cannot be used for initialization. The
reason is that the read-out only works if the ancilla qubit is initially
in a computational basis state; otherwise, it just projects onto the
subspace spanned by $\{\ket{0,0},\ket{1,1}\}$ or by $\{\ket{0,1},\ket{1,0}\}$.

In the following, we demonstrate that it is still possible to initialize
this scheme by taking a different perspective on how it encodes logical
qubits.  Therefore, we group each logical qubit with the ancilla above
(e.g.,\ the first two qubits in Fig.~\ref{fig:toric1}a), and encode the
new logical qubit in their parity -- note that this is what is really 
measured in the read-out.  The following calculations are 
 most conveniently carried out in a Bell basis where each
state is described as $\ket{s}_s\ket{l}_l$, where the $s$ qubit stores the
sign of the Bell state and the $l$ qubit the parity and thus encodes our logical qubit, i.e.
\begin{eqnarray}
\ket{s}_s\ket{0}_l &\leftrightarrow& \ket{0,0}+(-1)^s\ket{1,1}\\
\ket{s}_s\ket{1}_l &\leftrightarrow &\ket{0,1}+(-1)^s\ket{1,0}\ .
\end{eqnarray}
The circuit transforming between the above encoding and 
the qubits in correlation space is 
\begin{equation}
\label{eq:magictrafo}
\includegraphics[height=1.2cm]{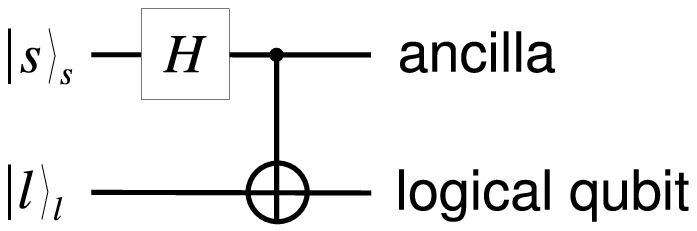}\ \ .
\end{equation}

Using this decoding, it is straightforward to investigate what happens in
the various steps of the MBQC scheme. Firstly, one can easily check that
by measuring two consecutive couplings of the qubit pair in the $X$ basis,
one prepares them in a maximally entangled state $\ket{0,0}+\ket{1,1}$ up
to Pauli errors, corresponding to $\ket{0}_s\ket{0}_l$ in the encoded
system. By pretending a Pauli $Z$ error on one of the qubits with
$p=1/2$, we effectively face the mixture $\ket{0,0}\bra{0,0}+
\ket{1,1}\bra{1,1}$, corresponding to
$\id_s\otimes\ket{0}\bra{0}_l$.

Since the transformation (\ref{eq:magictrafo}) is in the Clifford group,
Pauli errors remain Pauli errors in the encoded system. In the following,
we will check how the circuit acts on initial states
$\ket{\pm}_s\ket{0}_l$, where the sign can be different on each pair. 
As we will show, all of them give the same output statistics, and thus the
same holds for their mixture, i.e.\ the actual initial state. These
considerations are illustrated in Fig.~\ref{fig:toric2}, where we take the
circuits of Fig.~\ref{fig:toric1} and compose them with the decoding and
encoding circuits (boxed) in order to determine their action on the
encoded system.

Firstly, a $ZZ(\phi)$ gate on a pair gives a $Z$ rotation of the encoded
logical qubit, since the action of $ZZ(\phi)$ only depends on the parity
(Fig.~\ref{fig:toric2}a). The action of the second $ZZ$ rotation of
Fig.~\ref{fig:toric1}b which originally gave an $X$ rotation is shown in
Fig.~\ref{fig:toric2}b. The right hand side is obtained by using
$\mbox{\textsc{cnot}}=(\id\otimes H)\:CZ\:(\id\otimes H)$,
$H^2=\id$, the fact that diagonal operators commute,
and $(CZ)^2=\id$.
As we see from the simplified circuit, we obtain an $X$ rotation on the
upper logical qubit, but with the rotation direction determined by the
state of the $\ket{s}_s$ qubit below: While $\ket{+}_s$ results in a
rotation $R_x(\phi)$, the state $\ket{-}_s$ gives
\[
ZR_x(\phi)Z\propto R_x(-\phi)\ .
\]
Similarly, the circuit for the coupling of two logical qubits can be
simplified as in Fig.~\ref{fig:toric2}c: again, the coupling on the
logical qubits is
$\mathrm{Cpl}(\phi):=(H\otimes Z)ZZ(\phi)(H\otimes\id)$ or 
\[
(H\otimes ZX)ZZ(\phi)(H\otimes X)=
(Z\otimes\id)\mathrm{Cpl}(\phi)(Z\otimes\id)\ ,
\]
depending on whether the second $s$ qubit is $\ket{+}_s$ or $\ket{-}_s$. 

Therefore, the error introduced by the unknown state of each $s$ qubit
results in a $Z$ correction around each operation on the 
logical qubit above (note that we can assume this also for 
$Z$ rotations as they commute with 
the $Z$ correction).  Although the error itself is unknown and different
for each logical qubit, it is consistent within each qubit, as it is
always determined by the same ancilla. Thus, two subsequent $Z$ errors
cancel out, 
 and one remains only with one $Z$ correction on the logical
qubit at the beginning and one at the end of the sequence. The former has
no effect since the initial state is $\ket{0}_l$, while the latter has no
effect either since  the encoded logical qubit
is finally measured in the computational basis. Thus, the output
statistics for the circuit is independent of the initial state
$\ket{\pm}_s$ of the phase qubits, and one can equally well start from
their mixture $\id_s$ which completes the argument.

\subsubsection{Toric codes: second scheme}
\label{sec:toric2}

The second toric-code-like scheme  is based on a very different idea.
Therefore, observe that the $K_V$ tensor can be written as
\begin{equation}
	\begin{xy}
	*!C\xybox{\xymatrix@C=3mm@R=3mm{
	\ar[rd] & & \\
	& \xnode{K_V[s]}\ar[ru]\ar[rd] & \\
	\ar[ru] & & 
	}}\end{xy}=
	\begin{xy}
	*!C\xybox{\xymatrix@C=3mm@R=3mm{
	    \ar[rd]& & & & & \\
	    & \xnode{\textsc{copy}^\dagger}\ar@{-}[r]&
		\xnode{H}\ar@{-}[r]&\xnode{A[s]}\ar@{-}[r]
		&\xnode{\textsc{copy}}\ar[rd]\ar[ru] & \\
	    \ar[ru] & & & & & 
	}}\end{xy}
\end{equation}
where $\textsc{copy}$ is the copy gate $|0,0\rangle\langle0|+
|1,1\rangle\langle1|$, $H$ is the Hadamard gate (both with no physical
system associated to them), and $A$ the 1-D cluster projector, 
cf.\ Eqs.~(\ref{eqn:clusterMatrix0}) and (\ref{eqn:clusterMatrix1}). 
Thus, $K_V$ takes
two qubits in correlation space, projects them onto the 
$\{|0,0\rangle,|1,1\rangle\}$ subspace, implements the 1-D cluster map up to
a Hadamard, and duplicates the output to two qubits.  Concatenating these
tensors horizontally [this takes place in (\ref{eqn:nys:toriccode}) if all
$K_H$'s are measured in $Z$, and one neglects Pauli errors]
therefore implements a single logical qubit line, encoded in two
qubits in correlation space. By removing the Hadamard gate from $K_V$, we
obtain a 1-D cluster state encoded in two qubits which is thus capable of
implementing any one-qubit operation on the logical qubit; in particular,
this includes intialization and read-out. We thus define the tensor
\begin{equation}
	\begin{xy}
	*!C\xybox{\xymatrix@C=3mm@R=3mm{
	\ar[rd] & & \\
	& \xnode{\tilde K_V[s]}\ar[ru]\ar[rd] & \\
	\ar[ru] & & 
	}}\end{xy}=
	\begin{xy}
	*!C\xybox{\xymatrix@C=3mm@R=3mm{
	    \ar[rd]& & & &  \\
	    & \xnode{\textsc{copy}^\dagger}\ar@{-}[r]&
		\xnode{A[s]}\ar@{-}[r]
		&\xnode{\textsc{copy}}\ar[rd]\ar[ru] & \\
	    \ar[ru] & & & &  
	}}\end{xy}\ .
\end{equation}
  Then, the toric
code state (\ref{eqn:nys:toriccode}) with $K_V$ replaced by $\tilde K_V$ is
universal for MBQC: Initialization, one-qubit operations, and read-out
are done exacly as in the 1-D cluster state.
The logical qubits are decoupled up to $Z$ by-product operators in
correlation space by measuring the $K_H$ tensors in the $Z$ basis.
The $Z$ by-products in correlation space
correspond to $Z$ errors on the encoded logical
qubits and thus can again be dealt with as in the cluster. In order to
couple two logical qubits, we measure a $K_H$ tensor in the $Y$ basis and
obtain a
$ZZ$ controlled phase gate in correlation space, which translates to the
same gate on the logical
qubits. Note that this model has the additional feature that as 
as many controlled phases (between nearest neighbors) as desired can be
implemented simultaneously.

In the light of the discussion on the initialization of the first scheme,
one might see similarities between the two schemes, since in both cases
the information is effectively encoded in pairs of qubits. Note however
that in the first scheme, the information is stored in the parity of the
two
qubits, and the full $4$-dimensional space is being used; the reason for
this encoding came from the properties of the $K_H$ tensor used as a map in
horizontal direction. In contrast, the second scheme only populates the
$2$-dimensional even parity subspace, and the qubit is rather stored in two
copies of the same state; finally, the encoding is motivated by the
properties of the $K_V$ tensor as a map on correlation space in horizontal
direction.

\subsection{Weighted graph states}
\label{sec:weightedGraph}

In this section, we will consider instances of 
{\it weighted graph states}
\cite{Weighted,Survey} 
forming universal resources. 
To motivate the construction, recall that the 
cluster state can be
prepared by applying a controlled-phase gate
\begin{equation}\label{eqn:controlledPhase}
	P(\phi)=
	|0,0\rangle\langle 0,0 |
	+ |0,1\rangle\langle 0,1|+
	 |1,0\rangle\langle 1,0| + e^{i\phi} |1,1\rangle\langle 1,1|,
\end{equation}
with phase $\phi=\pi$ between any two nearest neighbors of a
two-dimensional lattice of qubits initially in the state $\ket+$.
If one wants to physically implement this operation using \emph{linear
optics} \cite{Linear}, one encounters the situation that the controlled
phase gate can be implemented only probabilistically, with the
probability of success decreasing as $\phi$ increases. It is hence
natural to ask whether one can build a universal resource using gates
$P(\phi), 0<\phi<\pi$, in order to minimize the probability of
failure\footnote{Alternative models with edges resulting 
	from commuting gates with non-maximally entangling
	power can possibly also be constructed by exploiting 
	ideas of non-local gates that are implemented with
	local operations and classical communication \cite{Non,Priv}.}

\begin{figure}
	\includegraphics[width=6.4cm]{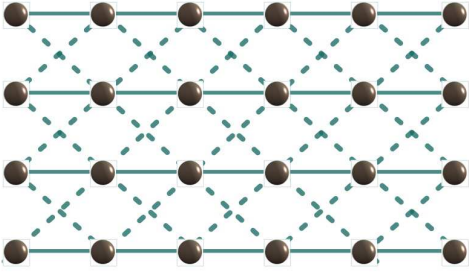}
	\caption{\label{fig:weighted}
		Weighted graph state as a universal resource. Solid lines 
		correspond to edges that have been entangled using phase
		gates with phase $\phi=\pi$, dotted lines correspond to 
		edges entangled with phase gates with $\phi=\pi/2$. This
		shows that one can replace some edges with 
		weakly entangled
		bonds.}
\end{figure}

%\begin{observation}[Quantum computing with weighted graph states] 
%A weighted graph state as depicted in Fig.\ \ref{fig:weighted} 
%is a universal resource for MBQC.
%\end{observation}

\subsubsection{Translationally invariant weighted graph states}

Expanding the discussion presented in Ref. \cite{shortOne}, we treat
the weighted graph state shown in Fig.\ \ref{fig:weighted}.  A tensor
network representation of these states can be derived along the same
lines as for the original cluster in Section \ref{sec:1dcluster}. Set
$\ket i:=2^{-1/2}(\ket0+i\ket1)$. 
%$\ket{\bar i}:=Z \ket i = 2^{-1/2}(\ket0-i\ket1)$. 
The relevant tensors are given by
\begin{eqnarray}
\label{eqn:triangleDef}  
\hspace{-1mm}
\begin{xy}
*!C\xybox{\xymatrix@C=3mm@R=1mm{
								&    & \\
 \ar[r]&\xnode{A[0]}\ar[lu]\ar[r]\ar[ru]& \\
 \ar[ru]&                            &\ar[lu]
}}
\end{xy}
&=&\ket +_{ru}\,\ket +_{lu}\,\ket+_r\bra 0_{ld}\bra 0_{rd}\bra 0_l, \\
\nonumber\\
\hspace{-1mm}
\begin{xy}
*!C\xybox{\xymatrix@C=3mm@R=1mm{
								&    & \\
 \ar[r]&\xnode{A[1]}\ar[lu]\ar[r]\ar[ru]& \\
 \ar[ru]&                            &\ar[lu]
}}
\end{xy}
&=&\ket i_{ru}\,\ket i_{lu}\,\ket-_r\bra 1_{ld}\bra 1_{rd}\bra 1_l. 
\end{eqnarray}
Indices are labeled $ru$ for ``right-up'' to $ld$ for
``left-down''.  The boundary conditions are $\ket 0$ for the
$ru,lu,r$-directions; $\ket+$ otherwise.

We will first describe how to realize isolated evolutions of single logical
qubits in the sense of Fig.\ \ref{fig:flow}. Again the strategy will
be to measure the sites of one horizontal line of the lattice in the
$X$-basis and all vertically adjacent systems in the $Z$-basis. The
analysis of the situation proceeds in perfect analogy to the one given
in Section \ref{sec:2dcluster}. One obtains
\begin{equation}\label{eqn:2dto1d}
	\begin{xy}
	*!C\xybox{\xymatrix@C=3mm@R=0mm{
	\xmnode{Z_{i-1,u}}&    &\xmnode{Z_{i+1,u}} \\
					\ar[r]	 &\xmnode{X_i}\ar@{-}[lu]\ar@{-}[ru]\ar[r]& \\
	\xmnode{Z_{i-1,d}}\ar@{-}[ru]&    &\xmnode{Z_{i+1,d}}\ar@{-}[lu] \\
	}}
	\end{xy}
	=H S^{2x_i+z_i},
\end{equation}
where 
\begin{equation}
	z_i=z_{i-1,u}+z_{i-1,d}+z_{i+1,u}+z_{i+1,d},
\end{equation}	 
and
	$S:= \operatorname{diag}(1,i)$
denotes the {\it $\pi/4$ gate}.

%\begin{lemma}[Matrices with finite order]\label{lem:by-products}
%	Let $A,B$ be matrices having finite order, 
%	i.e., there exists constants $n_{A/B}$ such that 
%	$A^{n_A}=B^{n_B}=\id$.  Every
%	element in the group generated by $A,B$ can be written as 
%	\begin{equation}
%	A B^{k_1}
%	A B^{k_2} A \dots A B^{k_n} 
%	\end{equation} for some $n\in \nn $
%	and $k_i \in \{0,1\}$.
%\end{lemma}

The operators $H$ and $S$ generate the 24-element single qubit
Clifford group. Following the approach of Section \ref{sec:aklt}, we
take this as the model's by-product group. 

Now choose some phase $\phi$. Re-doing the calculation which led to
Eq.\ (\ref{eqn:2dto1d}), where we now measure in the $\{\ket0\pm
e^{i\phi}\ket1 \}$-basis instead of $X$ on the central node, shows that the
evolution of the correlation space is given by $S(\phi)$, up to
by-products. In complete analogy to Section \ref{sec:aklt}, we see
that the model allows for the realization of arbitrary $SU(2)$
operations.

How to prepare the state of the correlation system for a single
horizontal line and how to read read it out has already been discussed
in Section \ref{sec:1dcluster}. Hence the only piece missing for
universal quantum computation is a single entangling two-qubit gate.

The schematics for a controlled-$Z$ gate between two horizontal lines
in the lattice are given below. We implicitly assume that all
adjacent sites not shown are measured in the $Z$-basis,
\begin{equation}
	\begin{xy}
	*!C\xybox{\xymatrix@C=5mm@R=1mm{
	\ar[r]&\xmnode{X}\ar@{-}[r]& \xmnode{X}\ar@{-}[r]& \xmnode{X}\ar[r]& \\
        &          &\xmnode{Y}\ar@{-}[lu]\ar@{-}[ru]\\
	\ar[r]&\xmnode{X}\ar@{-}[r]\ar@{-}[ru]&\xmnode{X}\ar@{-}[r]&\xmnode{X}\ar@{-}[lu]\ar[r]& 
	}}
	\end{xy}.
\end{equation}
The measurement scheme realizes a controlled-$Z$ gate, where the
correlation system of the lower line carries the control qubit and the
upper line the target qubit.

In detail one would proceed as follows: first one performs the
$X$-measurements on the sites shown and the $Z$-measurements on the
adjacent ones.  If any of these measurements yields the result
``$1$'', we apply a $Z$-measurement to the central site and restart
the procedure three sites to the right. This approach has been chosen
for convenience: it allows us to forget about possible phases
introduced by other measurement outcomes. Still, the ``correct''
result will occur after a finite expected number of steps, so the
overhead caused due to this simplification is only linear. It is also
not hard to see that most other outcomes can be 
compensated for -- so
for practical purposes the scheme could be vastly 
optimized.

Now assume that all measurements yielded ``$0$''. Then a
$Y$-measurement is performed on the central site, obtaining the result
$y$. As we did in Section \ref{sec:aklt2D}, we assume that
the (lower) control line is in the basis
state $\ket c$, for $c\in\{0,1\}$. The contraction of the lower-most three
tensors gives
\begin{eqnarray}
&&
\begin{xy}
*!C\xybox{\xymatrix@C=3mm@R=3mm{
								&&&\\
	\xnode{\ket c}\ar@{-}[r]&\xmnode{X}\ar[u]\ar@{-}[r]& \xmnode{X}\ar@{-}[r]&\xmnode{X}\ar[r]\ar[u]& \\
}}
\end{xy}
\\
\nonumber\\
&=& S^c \ket+_{lu}S^c \ket +_{ru} H\ket c_r, \nonumber
\end{eqnarray}
where as before $S=S(i)=\operatorname{diag}(1,i)$. 
We plug this result into the $A[Y]$ tensor:
\begin{eqnarray}
&&
\begin{xy}
*!C\xybox{\xymatrix@C=3mm@R=3mm{
								&&&\\
								&\xmnode{Y}\ar[lu]\ar[ru]\\
			 \xnode{S^c\ket+}\ar@{-}[ru] &                 & \xnode{S^c\ket+}\ar@{-}[lu]
}}
\end{xy}
\hspace{-5mm}
\\
\nonumber\\
&=&\ket+_{lu}\,\ket+_{ru} + (-1)^{c+y}i(S\otimes
S)\ket+_{lu}\,\ket+_{ru}. \nonumber
\end{eqnarray}
Lastly, for $x\in\{0,1\}$, 
\begin{eqnarray}
\begin{xy}
	*!C\xybox{\xymatrix@C=3mm@R=3mm{
	\ar[r]&\xmnode{X}\ar@{-}[r]& \xmnode{X}\ar@{-}[r]& \xmnode{X}\ar[r]& \\
				& \xnode{S^x\ket+}\ar@{-}[u] &                 & \xnode{S^x\ket+}\ar@{-}[u]
	}}
	\end{xy} = H Z^x.
\end{eqnarray}
Hence, the evolution on the upper line is 
\begin{equation}	
	H(\id + (-1)^{c+y} i Z) \propto H S Z^{y+c}, 
\end{equation}
equivalent to $Z^c$ up to
by-products. We arrive hence at the following conclusion:

\begin{observation}[Non-maximal entangling power]\label{obs:entanglement}
	Universal resouces may be prepared using commuting
	gates with non-maximal entangling power. 
\end{observation}

\begin{figure}
	\includegraphics[width=6cm]{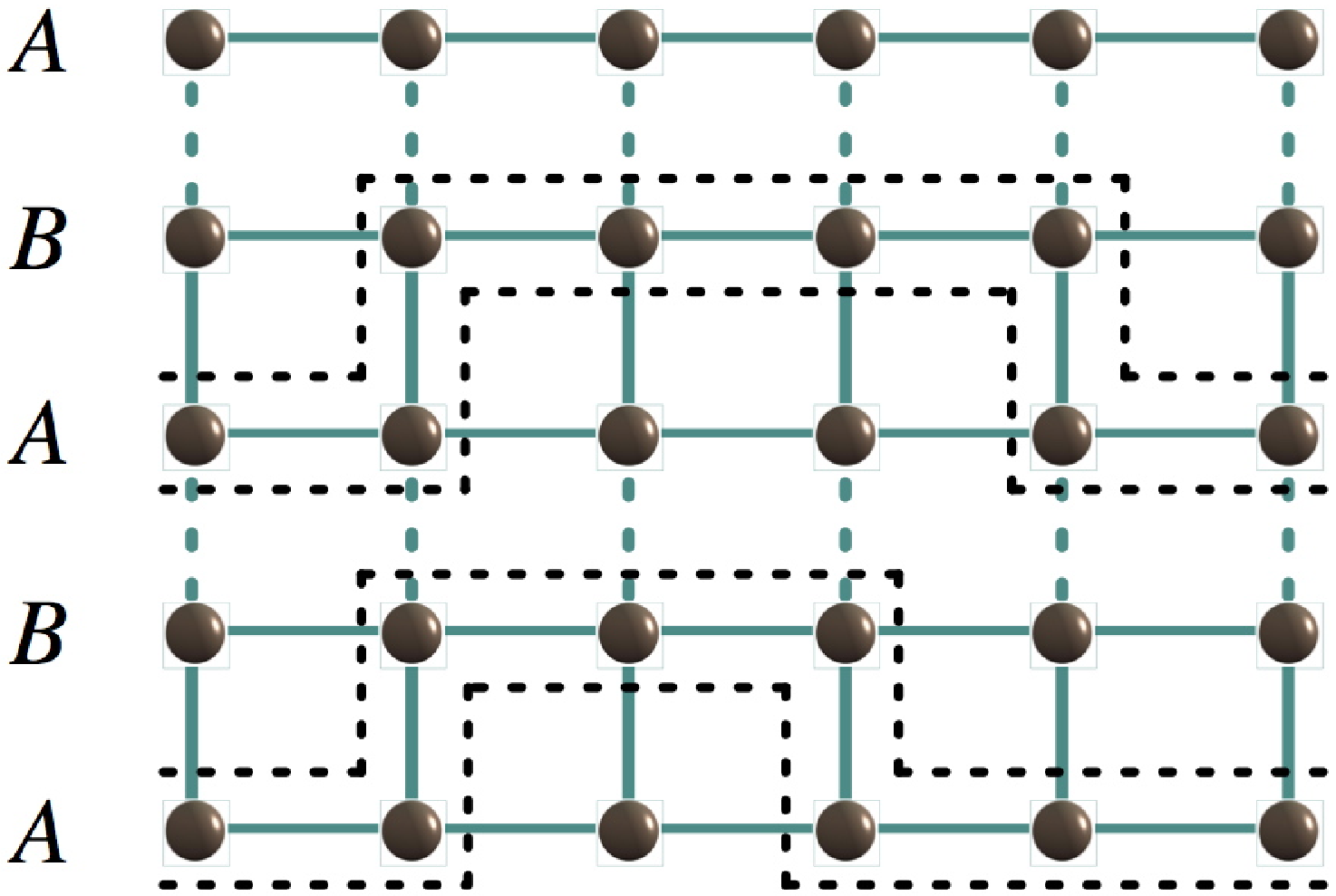}
	\caption{\label{fig:rerouting}
		Weighted graph state where the gate is achieved by appropriately
		bringing two wires together in a ``rerouting process''.}
\end{figure}

\subsubsection{Rerouting}

we will consider a second weighted graph state to exemplify yet
another novel ingredient that one can make use of in measurement-based
quantum computation: One can think of quantum information being
transported in the correlation system of some systems on the lattice
forming ``wires'', in a way that gates are realized by bringing the
``wires'' together. This is an element that is not present in the
original one-way computer. The subsequent example of a resource state
has not been chosen for its plausibility in the preparation in a
physical context, but in a way such that this idea of ``rerouting
quantum information'' can very transparently be explained, see Fig.\
\ref{fig:rerouting}.

The resource that we think about is defined by tensors that are
fully translationally invariant in one dimension, and 
has period two in the orthogonal 
dimension, 
\begin{equation}
	\begin{xy}
	*!C\xybox{\xymatrix@C=5mm@R=3mm{
	& \ar@{-}[d]& \ar@{-}[d]&\ar@{-}[d]& \ar@{-}[d]&\ar@{-}[d] & \\
	\ar@{-}[r]&\xnode{A}\ar@{-}[r]& 
	\xnode{A}\ar@{-}[r]&\xnode{A}\ar@{-}[r] &\xnode{A}\ar@{-}[r]&\xnode{A}\ar@{-}[r]& \\
	\ar@{-}[r]&\xnode{B}\ar@{-}[r]\ar@{-}[u]\ar@{-}[d]&\xnode{B}\ar@{-}[u]\ar@{-}[d]\ar@{-}[r]&\xnode{B}\ar@{-}[u]\ar@{-}[d]\ar@{-}[r]&
	  \xnode{B}\ar@{-}[u]\ar@{-}[d]\ar@{-}[r]&\xnode{B}\ar@{-}[u]\ar@{-}[d]\ar@{-}[r]&\\
	\ar@{-}[r]&\xnode{A}\ar@{-}[r]\ar@{-}[d]& \xnode{A}\ar@{-}[d]\ar@{-}[r]&\xnode{A}\ar@{-}[d]\ar@{-}[r] &\xnode{A}\ar@{-}[d]\ar@{-}[r]&\xnode{A}\ar@{-}[d]\ar@{-}[r]& \\
	& & & & & & }}
	\end{xy}.
\end{equation}
This is, we have two kinds of tensors: One set is given by
\begin{eqnarray}
	\label{eqn:rerouting0}
	\begin{xy}
	*!C\xybox{\xymatrix@C=3mm@R=3mm{
	 &&\\
	 \ar[r]&\xnode{B[0]}\ar[u]\ar[r]& \\
	 &\ar[u]&
	}}
	\end{xy}
	=\ket+_r\ket+_u\,\bra0_l\bra0_d, \\
	\label{eqn:rerouting1}
	\begin{xy}
	*!C\xybox{\xymatrix@C=3mm@R=3mm{
	 &&\\
	 \ar[r]&\xnode{B[1]}\ar[u]\ar[r]& \\
	 &\ar[u]&\
	}}
	\end{xy}
	=\ket-_r\ket i_u\,\bra1_l\bra 1_d
\end{eqnarray}
whereas the other one is nothing but the familiar one 
for a 2-D cluster state as in 
Eqs.\ (\ref{eqn:rerouting0}, \ref{eqn:rerouting1}),
with boundary conditions
\begin{equation}
	\ket L=\ket D =\ket+,
	\qquad
	\ket R=\ket U=\ket1.
\end{equation}	
The resulting state is hence again a weighted graph state, where
in one dimension every second edge is replaced by an edge
prepared using a gate with non-maximal entangling power.
Then, it is not difficult to see that, again with $x,z_r,z_u,z_d,z_l\in\{0,1\}$,
\begin{equation}
	\begin{xy}
	*!C\xybox{\xymatrix@C=3mm@R=3mm{
		&&&\\
	 	&\xnode{A[Z_u]}\\
	 	\ar[r]&\xnode{B[X]}\ar@{-}[u]\ar[r]&\\
	 	&\xnode{A[Z_d]}\ar@{-}[u]\\
	}}
	\end{xy} = HZ^{x+z_d}S^{z_u},
\end{equation}
and
\begin{equation}
	\begin{xy}
	*!C\xybox{\xymatrix@C=3mm@R=3mm{
		&&&\\
	 	&\xnode{B[Z_u]}\\
	 	\ar[r]&\xnode{A[X]}\ar@{-}[u]\ar[r]&\\
	 	&\xnode{B[Z_d]}\ar@{-}[u]\\
	  &\ar[u]&
	}}
	\end{xy} = HZ^{x+z_u}S^{z_d}.
\end{equation}
Similarly, we can consider several corner elements in this resource. We obtain
\begin{equation}
	\begin{xy}
	*!C\xybox{\xymatrix@C=3mm@R=3mm{
	 	& &\\
	 	\ar[r]&\xnode{A[X]}\ar[u]&\xnode{A[Z_r]}\ar@{-}[l]\\
	 	&\xnode{B[Z_d]}\ar@{-}[u]&\\
	}}
	\end{xy} = HZ^{x+z_d}S^{z_u},
\end{equation}
and similarly
\begin{eqnarray}
	\begin{xy}
	*!C\xybox{\xymatrix@C=3mm@R=3mm{
	 	 &\ar[d] &\\
	 	\xnode{A[Z_l]}\ar[r]&\xnode{A[X]}\ar[r]&\\
	 	&\xnode{B[Z_d]}\ar@{-}[u]&\\
	}}
	\end{xy} &=& (HSH)^{z_d} X^{z_l+x},\\[.2cm]
	\begin{xy}
	*!C\xybox{\xymatrix@C=3mm@R=3mm{
	 	 &\xnode{A[Z_u]} \ar@{-}[d]&\\
	 	\ar[r]&\xnode{B[X]}\ar[d]&\xnode{B[Z_r]}\ar@{-}[l]\\
	 	&&\\
	}}
	\end{xy} &=& Z^{x+z_r} S^{z_u},\\
	\begin{xy}
	*!C\xybox{\xymatrix@C=3mm@R=3mm{
	 	 &\xnode{A[Z_u]} \ar@{-}[d]&\\
	 	\xnode{B[Z_l]}\ar@{-}[r]&\xnode{B[X]}\ar[r]& \\
	 	&\ar[u]&\\
	}}
	\end{xy} &=& H Z^{x+z_u+z_l} (SZ)^{z_u},
\end{eqnarray}
where we have again made use of the convention that 
$x=0$ corresponds to $|+\rangle $ and 
$x=1$  to $|-\rangle $. We need one 
more ingredient to the scheme, this is 
\begin{eqnarray}
\begin{xy}
	*!C\xybox{\xymatrix@C=3mm@R=3mm{
		&&&\\
	 	\xnode{B[Z_l]}\ar@{-}[r]&\xnode{B[0]}\ar[u]\ar[r]&\\
	 	&\ar[u]&
	}}
	\end{xy} &=&|+\rangle_r |+\rangle_u\langle 0|_d,\\
	\begin{xy}
	*!C\xybox{\xymatrix@C=3mm@R=3mm{
		&&&\\
	 	\xnode{B[Z_l]}\ar@{-}[r]&\xnode{B[1]}\ar[u]\ar[r]&\\
	 	&\ar[u]&
	}}
	\end{xy} &=&|-\rangle_r |i\rangle_u\langle 1|_d,
\end{eqnarray}
and
\begin{eqnarray}
\begin{xy}
	*!C\xybox{\xymatrix@C=3mm@R=3mm{
		&&&\\
	 	\ar[r]&\xnode{A[0]}\ar@{-}[r]\ar[u]&\xnode{A[Z_r]}&\\
	 	&\ar[u]& &
	}}
	\end{xy} &=& |+\rangle_u\langle 0|_l \langle 0|_d,\\
	\begin{xy}
	*!C\xybox{\xymatrix@C=3mm@R=3mm{
		&&&\\
	 	\ar[r]&\xnode{A[1]}\ar@{-}[r]\ar[u]&\xnode{A[Z_r]}&\\
	 	&\ar[u]& &
	}}
	\end{xy} &=& (-1)^{z_r}|-\rangle_u\langle 1|_l \langle 1|_d.
\end{eqnarray}
Putting these ingredients, and following an argument similar to the
last subsection, we find that up to Clifford group by-products, we can
transport along the horizontal lines for both kinds of local tensors.
We can also use the corner pieces to reroute as depicted in Fig.\
\ref{fig:rerouting}, and bring routes together forming a ``gate''
imprinted in the lattice, actually, a controlled-$S$ gate.  

It should be noted that it is not obviously possible to faithfully
transport one qubit of information vertically through the resource.
Loosely speaking, the entanglement between a site of type B and the
site of type A directly above it is non-maximal (this is indicated by
dotted lines in Fig.\ \ref{fig:rerouting}). Interestingly, one can
still perform a (non-maximally entangling) non-local gate over this
connection.

\begin{observation}[Rerouting] 
	Gates in measurement-based quantum computation can be achieved by
	means of appropriate routing of quantum information in the
	lattice.
\end{observation}

\subsection{A qubit resource with non-vanishing correlation functions}

We will very briefly sketch a matrix product state on a 1-D chain of
qubits, which i) exhibits non-vanishing two-point correlation
functions, ii) allows for any unitary to be realized in its
correlation system and iii) can be coupled to a universal 2-D resource
in a way very similar to the AKLT-type example (Section
\ref{sec:aklt}). The discussion will be somewhat superficial --
however, given the extensive discussion of other models above, the
reader should have no problems filling in the details.

Choose an integer $m>2$ and define 
\begin{equation}	
	G:=\exp(i \pi/m X). 
\end{equation}
Up to a constant, $G$ is a $m$-th root of $X$. The state is defined by
the following relations:
\begin{eqnarray}\label{eqn:gstate}
	\xymatrix@C=4mm{
		\ar[r]&\xnode{A[s]}\ar[r]&
	}&=& \ket s_r \bra s_l G,
\end{eqnarray}
and
\begin{equation}
	\ket L = G^\dagger \ket +,\quad \ket R = \ket +. \label{eqn:gstate2}
\end{equation}

%\begin{observation}[Quantum computing with long-range correlations] 
%Exponentially decaying but non-vanishing correlation functions -- as in the
%AKLT-type model and in the one of 
%Eqs.\ (\ref{eqn:gstate},\ref{eqn:gstate2}) -- do not constitute an 
%obstacle to universal quantum computation.
%\end{observation}

The two-point correlation functions for measurements on this state
never vanish completely. Indeed, in Appendix \ref{sec:correlations} it
will be shown that 
\begin{equation}
	\langle Z_i Z_{i+k}\rangle - \langle Z_i\rangle\>\langle
	Z_{i+k}\rangle = 2 \xi^k,
\end{equation}
where $\xi:=2\sin^2(\pi/m)-1$.

For $X$-measurements, we find
\begin{eqnarray}\label{eqn:phiMeas1}
	\xminid{\ar[r]&\xmnode{X}\ar[r]&} 
	&=& Z^x G
\end{eqnarray}
%	\xminid{\ar[r]&\xnode{A[\bar\phi]}\ar[r]&} 
%	&=& ZS(\phi) G,\label{eqn:phiMeas2}
%\end{eqnarray}
Pursuing the strategy introduced in Section \ref{sec:randomness}, we
set the by-product group to $\mathcal{B}=\langle Z, G\rangle$, so the
group generated by $Z$ and $G$. One can easily verify that
$\mathcal{B}$ is indeed a finite group, equivalent to the
\emph{dihedral group} of order $2m$.

It is now straight-forward to check that i) measurements in the
computational basis can be used for preparation and read-out (as in
Section \ref{sec:1dcluster}), ii) general local unitaries can be
realized by means of measurements in the equatorial plane of the Bloch
sphere (as in Section \ref{sec:1daklt}) and iii) a 2-D resource is
obtainable in a fashion similar to the one presented in Section
\ref{sec:2daklt}. With similar methods, one can also find
qubit resource states that have a local entropy smaller than unity.

\subsection{Percolation ideas to make use of imperfect resources}

For completeness, we mention yet another kind of resource: This is an
imperfect cluster state where some edges are missing. Such a setting
is clearly relevant in a number of physical situations: If the
underlying quantum gates building up the cluster state are
fundamentally probabilistic, such as in linear optical architectures,
then one very naturally arrives at this situation when one aims at
minimizing the need for feed-forward. A similar situation is
encountered in cold atoms in optical lattices, when in a Mott state
exhibiting hole defects some atoms are missing. We do not present
details of such arguments, which have been considered in Ref.\
\cite{Percolation}, based on ideas of {\it edge percolation} and
renormalization \cite{Grimmett}.  We merely state the result for
completeness. Note also
that results that may be similar to these ones have been 
announced in Ref.\ \cite{maarten1}.

We consider the setting where one starts from a 2-D or 3-D cubic
lattice of size $n\times n$. Two neighboring vertices on the lattice
are connected with an edge with probability $p$. The stochastic
variables deciding whether or not an edge is present are assumed to be
uncorrelated. If $p>p_2=1/2$ holds, then it is not difficult to see
that one can extract a 2-D renormalized lattice of smaller size: This
means that one can find a function $n\mapsto m(n)$, such that one
arrives at a cubic $m(n)\times m(n)$ array almost certainly as
$n\rightarrow\infty$, with the following property:
Within each of the elements of this array, there is a central
site that is connected to the central site of the neighboring
array. Since all the additional sites can be removed by means
of $Z$-measurements, we can treat this resource effectively
as a 2-D cluster state of dimension $m(n)\times m(n)$,
and refer to this as a {\it perfect sublattice}. This state will
not necessarily be exactly a cluster state, as it may contain
vertices having a vertex degree of three, but which will nevertheless
function as a graph state resource just as the cluster state
does (for details, see Ref.\ \cite{Percolation}).
Also, $n/m(n)$ is arbitrarily close to being
linear in $n$ asymptotically.  However, an even stronger statement
holds: 

\begin{observation}[Percolation] Whenever $p>p_3= 0.249$, 
for any $\varepsilon>0$, one can find a function $n\mapsto m(n)$ with the following property:
Starting from a sublattice of a 3-D cubic lattice of size $n\times n\times 2n/m(n)$, one can 
almost certainly prepare a perfect sublattice of 
size $m(n)\times m(n)$. The asymptotic behavior
of $m$ can be chosen to satisfy 
\begin{equation}
	n/m(n) = O(n^\varepsilon).
\end{equation}	
\end{observation}

\begin{figure}
	\includegraphics[width=4.5cm]{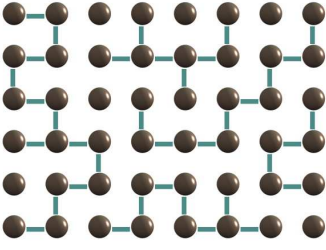}
	\caption{\label{fig:percolation}
		Cubic lattice of a graph state corresponding to the situation where
		some edges are missing in a cluster state. If the probability $p$
		of having an edge is sufficiently high the 
		processes independent,
		then a renormalized perfect sublattice can be 
		found almost 
		certainly, giving rise
		to a cluster state of smaller size. 
		If $p>p_2=1/2$, where $p_2$ is
		the percolation threshold
		for edge percolation in 2-D cubic lattices, 
		then a renormalized lattice can be found 
		almost certainly. Interestingly,
		even if $1/2>p>p_3$, $p_3= 0.249$ 
		denoting the percolation threshold in 3-D,
		one can almost certainly construct a perfect
		sublattice,
		using an overhead 
		that is arbitrarily close to being quadratic.}
\end{figure}

That is, with an overhead that is arbitrarily close to the optimal
scaling, one can obtain a perfect resource state out of an imperfect
one, even if one is merely above the percolation threshold for a
three-dimensional lattice, and not only for the two-dimensional
lattice, see Fig.\ \ref{fig:percolation}. The latter argument is
technically more involved than the former, for details, see Ref.\
\cite{Percolation}. This shows, 
however, with 
methods unrelated to the
ones considered primarily in the present work, that also 
random
aspects in the resource as such can be dealt 
with. 

\section{One-way computation using encoded systems}

In the final section of this work, we will show that one can find
resource states for MBQC that differ substantially from the cluster in
various entanglement properties. This will be done by encoding
each system of a resource into several physical particles. We will
not develop any new computational models and make no use of the
computational tensor network formalism introduced before. The study of
encoded resource states was initiated in Ref.\ \cite{shortOne} and
later pursued more systematically in Ref.\ \cite{InnsbruckLong}.

More concretely, the following statements will be proved:

\begin{observation}[Resources with weak capabilities for
state preparation]
There exists a family of universal resource states such that 
\begin{itemize}
	\item
	The local entropy of entanglement is arbitrarily small,

	\item
	The localizable entanglement is arbitrarily small
\end{itemize}
and, more strongly,
\begin{itemize}
	\item
	The probability of succeeding in distilling a maximally entangled
	pair out of the resource is arbitrarily small, even if one does not
	a priori fix the two sites between which the pair will be
	established.
\end{itemize}
In particular, the resource cannot be used as a state preparator.
\end{observation}

We start from a cluster state vector
on $n\times n$ systems, denoted by
$\ket{Cl_{n\times n}}$, referred to as logical qubits. 
As in Ref.\ \cite{shortOne},
we want to ``dilute'' the cluster state, i.e.
encode it into a larger system, by means of invoking 
the codewords
\begin{equation}
	\ket{\tilde 0}:=\ket 0^{\otimes k}, \,\,
	\ket{\tilde 1}:=\ket{W_k} 
\end{equation}	
	for some
parameter $k$. The argument relies only on the choice of 
 $\ket{W_k} $ as a code word in that we focus on its
implications on the localizable entanglement, and for that argument,
the state vector  $\ket{W_k} $ has the desired 
properties of small local entropy and permutation invariance.
However, for encoded one-way computation to be 
possible, any state vector orthogonal to  $\ket 0^{\otimes k}$ may be 
taken, compare also Ref.\ \cite{InnsbruckLong}.
Every qubit of the cluster is subjected to the encoding
operation 
\begin{equation}	
	V:=\ket{\tilde 0}\bra0+\ket{\tilde 1}\bra1
\end{equation}	
yielding the
\emph{diluted cluster} $\ket{{\cal D}_{n,k}}$.  A set of physical
qubits corresponding to one cluster bit will be called a 
\emph{block}.  As before, by 
a \emph{local measurement scheme} we mean a 
sequence of adaptive
local projective measurements, local to the physical systems. 

Let us first show again in more detail that such an encoding
constitutes no obstacle to universal quantum computation. 
Each of the code words is orthogonal, and for computation to be 
possible, we need to do local dichotomic measurements in the
logical space. 
By Ref.\ \cite{Walgate}, any two pure orthogonal
multi-partite states on $k$ qubits
can be deterministically distinguished 
using LOCC. By making use of the construction 
of Ref.\ \cite{Walgate}, this can be done by an appropriate 
ordered sequence of adapted projective measurements 
$\pi_1\otimes \dots \otimes \pi_k$ on the sites
of each codeword, giving rise to an 
arbitrary projective dichotomic measurement 
with Kraus operators 
\begin{equation}
	A_1 := |\psi\rangle\langle\psi|,\,\,
	A_2 := |\psi^\bot \rangle\langle\psi^\bot | = 
	\id - |\psi\rangle\langle\psi|
\end{equation}	
in the logical space, $|\psi\rangle =\alpha |0\rangle + \beta |1\rangle$
and $|\psi^\bot\rangle =- \beta^\ast  |0\rangle + \alpha^\ast | 1\rangle$.
%Clearly, the encoding amounts to
%\begin{eqnarray}
%	|\psi\rangle|\phi_1\rangle + |\psi^\bot \rangle|\phi_2\rangle
%	&\mapsto &
%	\ket{\tilde 0}(\alpha |\tilde \phi_1\rangle -  \beta^\ast |\tilde\phi_2 %\rangle)
%	\nonumber\\
%	&+& \ket{\tilde 1}(\beta |\tilde \phi_1\rangle + \alpha^\ast |\tilde%%\phi_2 \rangle),
%\end{eqnarray}
%$ |\tilde \phi_k\rangle$ being the encoded state vectors of $ | \phi_k\rangle$.
Hence,  
one can translate any single-site measurement on a cluster state
into an LOCC protocol for the encoded cluster. This shows that
$\ket\Psi$ is universal for deterministic MBC. This is the argument of 
Ref.\ \cite{shortOne} (see 
also Ref.\ \cite{InnsbruckLong} for a more 
detailed and extensive discussion on one-way computing 
based on encoded systems).

In the following we are going to show in more detail 
that despite this property, we are heavily
restricted to use this resource to prepare states with a significant
amount of entanglement between two constituents. In fact, we can not
even distill a perfect maximally entangled qubit pair beyond any
given probability of success. This means that these states are universal
resources, but on the level of physical systems utterly useless for
state preparation. The given resource is, needless to say, not meant as a
particularly feasible resource. Instead, we aim at highlighting to what extent as
such the entanglement properties can be relaxed, giving a guideline to more
general settings.  

Note first that the localizable entanglement $E_L$
	in these resources can easily be shown
	to be arbitrarily small: The 
	entropy for a measurement in the computational basis reads
	$H_b(3/(4k+2))$, where $H_b:[0,1]\rightarrow[0,1]$ 
	is the standard 
	binary entropy function.
	Using the concavity of the entropy function, we find 
	\begin{equation}	
		E_L(\ket{{\cal D}_{n,k}}\bra{{\cal D}_{n,k}}) \leq
		H_b(3/(4k+2)),
	\end{equation}
	such that $\lim_{k\rightarrow\infty} E_L(\ket{{\cal D}_{n,k}}\bra{{\cal D}_{n,k}}) =0$.
	This means that for two fixed sites, the rate at which one can distill maximally
	entangled pairs by performing measurements on the remaining systems is arbitrarily
	small.

This can be seen as follows. We will aim at preparing a maximally
entangled state between any two constituents of two different blocks. It is easy
to see that within the same block, the probability of success can be made
arbitrarily small. We hence look at a LOCC distillation scheme, a
{\it measurement-based scheme}, taking the input
$\rho$ and producing outputs
\begin{equation}
	\rho\mapsto K_j \rho K_j^\dagger
\end{equation}
with probability $p_j= \tr(K_j \rho K_j^\dagger)$, $j=1,\dots, J$. This corresponds to a LOCC
procedure, where each of the measurements may depend on all outcomes of the previous
local measurements. 
Let us assume that outcomes labeled $1,\dots, S$ for some 
$S\leq J$ are successful in
distilling a maximally entangled state.  

We start by exploiting the permutation symmetry of the code words.
	Choose a block $i$ of $\ket{{\cal D}_{n,k}}$. Assume there exists a
	measurement-based scheme with the property that
		with probability $p$, the scheme will leave \emph{at least one} system
		of block $i$ in a state of maximal local entropy.
	Then there exists a scheme such that
		with probability $p$, the scheme will leave \emph{the first} system
		of block $i$ in a state of maximal local entropy.
	At some point of time the scheme is going to perform the first
	measurement on the $i$-th block. Because of permutation invariance,
	we may assume that it does so on the $k$-th system of the block.
	The remaining state is still invariant under permutations
	of the first $k-1$ systems. Hence there is no loss of
	generality in assuming that the next measurement on the $i$-th block
	will be performed on the $k-1$-st system. If
	the local entropy of any of the unmeasured systems is now maximal, then
	the same will be true for the first one -- once again, by
	permutation invariance.
 
 	Also, it is easy to see that the probability $p$
	that a measurement-based scheme
	will leave any system of block $i$ in a locally maximally mixed
	state is bounded from above by
	\begin{equation}
		p<2/k.
	\end{equation}	
	Let $p_1$ be the initial probability of obtaining the outcome $\ket1$ for a $Z$
	measurement on this qubit, 
	$p_1=|\langle 1 | {{\cal D}_{n,k}}\rangle|^2$.
	Clearly, 
	\begin{equation}
		p_1< 1/k.
	\end{equation}
	We consider now a local scheme 
	potentially acting on all qubits except
	this distinguished one, with branches labeled
	$j=1,...,J$, aiming at preparing this qubit in a 
	maximally mixed state.
	Let $p_s$ be the probability of the 
	qubit
	ending up in a locally maximally mixed state.
	In case of success, so in case of the preparation of a 
	locally maximally entangled state,
	we have that 	
	$p_1(s) = 1/2$, 
		in case of failure
		$p_1(f) \geq 0$. 
	Combining these inequalities, we get
	\begin{eqnarray}\label{eqn:ineq}
		1/k > p_1 
		=  p_s p_1(s) + (1-p_s) p_1(f)  
		=  p_s/2.
	\end{eqnarray}
	We can hence show that there exists a family of universal resource states
	such that the probability that a local measurement scheme can
	prepare a maximally entangled qubit pair (up to l.u. equivalence) 
	out of any element of
	that family is strictly smaller than $\varepsilon>0$.
 
	Let $p_i$ be the probability that a site of 
	block $i$ will end up as
	a part of a maximally entangled pair.  
	This means that when we fix the procedure, and
	label as before all sequences of measurement outcomes
	with $j=1,...,J$, one  does not perform measurements
	on all constituents. Let $I$ denote the index set
	labeling the cases where somewhere on the lattice
	a maximally entangled pair
	appears, so the probability $p$ for this to happen is bounded
	from above by
	\begin{equation}
		p\leq \sum_{i\in I} 
		p_i.
	 \end{equation}
	 According to the above bound,	 
	 $p_i<2/k$, giving
	a strict upper bound of $p \leq
	2 n^2/k$ for the overall probability of
	success. The family 
	\begin{equation}	
		\ket{\Psi_n}:=\ket{{\cal D}_{n,k(n)}},
	\end{equation}	
	for $k(n) :=2 \varepsilon^{-1} n^2$ is clearly universal, involves 
	only a linear overhead as compared to the original cluster 
	state and satisfies the assumptions advertised above. 
			
\section{Conclusions}
	
	In this work, we have shown how to construct a plethora of novel
	models for measurement-based quantum computation. 
	Our methods were taken from many-body theory. 	
	The new models for quantum computation follow the
	paradigm of locally measuring single sites -- and 
	hence abandoning
	any need for unitary control during the computation. Other than
	that, however, they can be quite different from the 
	one-way model.
	We have found models where the randomness is 
	compensated in a novel
	manner, the length of the computation can be random, gates are
	performed by routing flows of quantum information towards one
	another, and logical information may be encoded in many 
	correlation
	systems at the same time.  What is more, the resource states 
	can in
	fact be radically different from the cluster states, in that
	they may display correlations as typical in ground states, can be 	weakly
	entangled. 
	%or be variants of Kitaev's toric code states. 
	A number of
	properties of resource states that we found 
	reasonable to assume to
	be necessary for a state to form a universal resource could be
	eventually relaxed. So after all, it seems that 
	much less is needed
	for measurement-based quantum computation than 
	one could reasonably
	have anticipated. 
	This new degree of flexibility may well pave the
	way towards tailoring computational model towards 
	many-body states
	that are particularly feasible to prepare, rather than trying to
	experimentally realize a specific model.
	
\section{Acknowledgements}

This work has benefited from 
fruitful discussions with  
a number of people, including 
K.\ Audenaert,
I.\ Bloch, 
H.-J.\ Briegel, 
J.I.\ Cirac,
C.\ Dawson, 
W.\ D{\"u}r,
D.\ Leung,
A.\ Miyake,
M.\ Van den Nest, 
F.\ Verstraete, 
M.B.\ Plenio,
T.\ Rudolph, 
M.M.\ Wolf, and 
A.\ Zeilinger.
It has been supported by the EU (QAP, QOVAQIAL), the 
Elite-Netzwerk Bayern,
the EPSRC, the QIP-IRC, Microsoft Research,
and the EURYI Award Scheme.

\section{Appendix}

\subsection{Computing correlations functions}
\label{sec:correlations}

What is the value of the two-point correlation function 
$\langle Z_i Z_{i+k} \rangle-\langle Z_i\rangle\,\langle
Z_{i+k}\rangle$? In this
work, we have only introduced the behavior of the correlation system
when subject to a local measurement of a rank-one observable. However,
in order to evaluate the correlation function, we need  ``measure the
identity'' on the intermediate systems or, equivalently, trace them
out. Without going into the general theory \cite{FCS}, we just state that
tracing out a system will cause the completely positive map
\begin{equation}
	\Phi: \rho \mapsto \sum_i A[i]\rho A[i]^\dagger
\end{equation}
to act on the correlation system. 

For the cluster state, using the fact that the bases $\{\ket0,\ket1\}$
and $\{\ket+,\ket-\}$ are unbiased, we can easily show that $\Phi^2$ is
the completely depolarizing channel, sending any $\rho$ to $2^{-1}\id$.
This causes any correlation function to vanish for $k>2$. How does the
situation look like for the state vector defined by Eq.\
(\ref{eqn:gstate})? We compute:
\begin{equation}
	\Phi: \rho \mapsto 
	\sum_{s=0,1}
	\tr(\rho\, G\ket s\bra s G^\dagger)\, \ket0\bra0,
%	+ 
%	\tr(\rho G\ket1\bra1 G^\dagger) \ket1\bra1,
\end{equation}
so for $s\in\{0,1\}$: 
\begin{equation}
	\Phi(\ket s\bra s)=p \ket s\bra s + (1-p)\ket{\bar s}\bra{\bar s}
\end{equation}
where $\bar 0:=1, \bar 1:=0$ and $p:=|\bra 0 G \ket 0|^2=\sin^2(\pi/m)$.
In other words: when acting on the computational basis, $\Phi$
implements a simple two-state Markov process, which remains in the
same state with probability $p$ and switches its state with
probability $(1-p)$. Now, $\langle Z_i Z_{i+k}\rangle$
equals $+2$ if an even number of state changes occurred and $-2$ if
that number is odd. So for the expectation value we find
\begin{eqnarray}
	\langle Z_i Z_{i+k}\rangle 
	&=& 2 \sum_{l=0}^{k+1} {k \choose l} p^{k-l}(1-p)^k (-1)^k \\ 
	&=& 2 (2p-1)^k = 2 (2\sin^2(\pi/m)-1)^k.\nonumber
\end{eqnarray}

\subsection{Hamiltonian of the AKLT-type state}
\label{sec:akltAppendix}

In Section \ref{sec:aklt} we discussed an AKLT-type matrix product
state.
It was claimed that the state constitutes the unique
ground-state of a spin-1 nearest neighbor frustration free
gapped Hamiltonian. It must be noted that in this work, we have not
introduced the technical tools needed to cope with boundary effects at
the end of the chain. There are at least three ways to make the above
statement rigorous: a) treat the statement as being valid
asymptotically in the limit of large chains, b) work directly with
infinite-volume states \cite{FCS}, or c) look at sufficiently large
rings with periodic boundary conditions \cite{david}.  Once one
chooses one of the options outlined above, the proof of this fact
proceeds along the same lines as the one of the original AKLT state,
as presented in Example 7 of Ref.\ \cite{FCS} (see also Ref.\ 
\cite{david}).
Indeed, using the notions of Refs.\ \cite{FCS,david} one verifies 
that
\begin{eqnarray}
	&\Gamma_2: {\cal B}(\cc^2) \to \cc^2\otimes \cc^2,&\\
	&B \mapsto \sum_{i_1, i_2=1}^3 \tr(B A[i_1] A[i_2]) 
	\ket{i_1,i_2}&
\end{eqnarray}
is injective. Further, if ${\cal G}_2:=\operatorname{range} \Gamma_2$,
it is checked by direct computation that $\dim({\cal G}_2\otimes\id
\cap \id\otimes {\cal G}_2)=\dim {\cal G}_2$. All claims follow as
detailed in Refs.\ \cite{FCS,david}.

In particular, let $h$ be a positive operator supported on the vector
space spanned by:
\begin{eqnarray}
&&\{ |1,1 \rangle, |2,2\rangle, 
-(1/4) |0,0\rangle+ |1,2\rangle+ |2,1\rangle, \\
&& -(1/\sqrt8) |0,0\rangle+ |0,2\rangle+ |2,0\rangle, \nonumber \\
&& -(1/\sqrt8) |0,0\rangle+ |0,1\rangle+ |1,0\rangle\}.\nonumber
\end{eqnarray}
Set $H:=\sum_i \tau_i(h)$, where $\tau_i$ translates its argument $i$
sites along the chain. Then $H$ is a non-degenerate, gapped,
frustration free, nearest neighbor Hamiltonian (called \emph{parent
Hamiltonian} in Ref.\ \cite{david}), whose energy is minimized by the
state at hand.


\begin{thebibliography}{99}

\bibitem{shortOne}
	D.\ Gross and J.\ Eisert, Phys.\ Rev.\ Lett.\ {\bf 98},
	220503  (2007).

\bibitem{Oneway}
        R.\ Raussendorf and H.-J.\ Briegel,
        Phys.\ Rev.\ Lett.\ {\bf 86},  5188 (2001).

\bibitem{OnewayCompModel}
	R.\ Raussendorf and H.-J.\ Briegel, 
	Quant.\ Inf.\ Comp.\ {\bf 6}, 433 (2002).

\bibitem{Cluster}
       H.-J.\ Briegel and R.\ Raussendorf,
        Phys.\ Rev.\ Lett.\ {\bf 86}, 910 (2001).
       
\bibitem{Survey}        
        M.\ Hein, W.\ D{\"u}r, J.\ Eisert,
        R.\ Raussendorf, M.\ Van den Nest, and
        H.-J.\ Briegel,
        quant-ph/0602096.

\bibitem{CompSurvey}        
	M.A.\ Nielsen, quant-ph/0504097;
         D.E.\ Browne and H.-J.\ Briegel, quant-ph/0603226.
         
\bibitem{GS}
        M.\ Hein, J.\ Eisert, and H.-J.\ Briegel,
        Phys.\ Rev.\ A {\bf 69}, 062311 (2004);
        R.\ Raussendorf, D.E.\ Browne, and H.-J.\ Briegel,
        ibid.\ {\bf 68}, 022312 (2003);
        D.\ Schlingemann and R.F.\ Werner,
        ibid.\ {\bf 65}, 012308 (2002).
        
\bibitem{Greiner}
	O.\ Mandel, M.\ Greiner, A.\ Widera, T.\ Rom, T.W.\ H{\"a}nsch, 
	and I.\ Bloch, Nature {\bf 425}, 937 (2003).
		
\bibitem{Plenio}
	M.J.\ Hartmann, F.G.S.L.\ Brandao, M.B.\ Plenio,
	Nature Physics {\bf  2}, 855 (2006);
	 A.D.\ Greentree, C.\ Tahan, J.H.\ Cole, and L.C. L.\ 
	 Hollenberg, ibid.\ {\bf 2}, 856 (2006).
			
\bibitem{Atoms}
	C.\ Cabrillo, J.I.\ Cirac, P.\ Garcia-Ferndandez, and 
	P.\ Zoller, Phys.\ Rev.\ A {\bf 59}, 1025 (1999);
	D.E.\ Browne, M.B.\ Plenio, and S.F.\ Huelga,
	Phys.\ Rev.\ Lett.\ {\bf 91}, 067901 (2003).

\bibitem{Zeilinger}
	 P.\ Walther, K.J.\ Resch, T.\ Rudolph, E.\ Schenck, 
	 H.\ Weinfurter, V.\ Vedral, M.\ Aspelmeyer, and A.\ 
	 Zeilinger, Nature {\bf 434}, 169 (2005).
	 
\bibitem{ClusterOptical}	 
	 D.E.\ Browne and T.\ Rudolph,
	 Phys.\ Rev.\ Lett.\ {\bf 95}, 010501 (2005).

\bibitem{ProbaLin}
	 D.\ Gross, K.\ Kieling, and J.\ Eisert,
	 Phys.\ Rev.\ A {\bf 74}, 042343 (2006);
	 K.\ Kieling, D.\ Gross, and J.\ Eisert;
	 J.\ Opt.\ Soc.\ Am.\ {\bf B 24}(2), 184 (2007).

\bibitem{Traps}
	 H.\ Haeffner et al., Nature {\bf 438}, 643(2005).
	 
\bibitem{vidal}
	 G.\ Vidal, Phys.\ Rev.\ Lett.\ {\bf 91}, 147902 (2003); R.\ Jozsa,
	 quant-ph/0603163; I.\ Markov and Y.\ Shi, 
	 quant-ph/0511069; Y.-Y.\
	 Shi, L.-M.\ Duan, and G.\ Vidal, Phys.\ Rev.\ A {\bf 74}, 022320
	 (2006). 
	 
\bibitem{maarten2}
	M.\ Van den Nest, W.\ D{\"u}r, G.\ Vidal, and H.J.\
	 Briegel, Phys.\ Rev.\ A {\bf 75}, 012337 (2007).

\bibitem{teleporters}
	D.\ Gottesman and I.L.\ Chuang, Nature {\bf 402}, 390 (1999);
	M.A.\ Nielsen, Phys.\ Lett.\ A. {\bf 308}, 96 (2003);
	D.W.\ Leung, quant-ph/0111122.

\bibitem{teleportersEquiv}
	P.\ Aliferis and D.W.\ Leung, Phys.\ Rev.\ A {\bf 70}, 062314
	(2004);
	A.M.\ Childs, D.W.\ Leung, and M.A.\ Nielsen, quant-ph/0404132
	(2004); P.\ Jorrand and S.\ Perdrix, quant-ph/0404125 (2004).
         
\bibitem{Jozsa}
	R.\ Jozsa, quant-ph/0508124.

\bibitem{OneWayPEPS}		
	F.\ Verstraete and J.I.\ Cirac,
	Phys.\ Rev.\ A {\bf 70}, 060302(R) (2004).		

\bibitem{aklt}	
	I.\ Affleck, T.\ Kennedy, E.H.\ Lieb, and H.\ 
	Tasaki, ibid.\ {\bf 59}, 799 (1987).

\bibitem{elham}
	V.\ Danos, E.\ Kashefi, and P.\ Panangaden, quant-ph/0704.1263
	(2007).

\bibitem{Vedral}
	 M.S.\ Tame, M.\ Paternostro, M.S.\ Kim, and 
	 V.\ Vedral, Phys.\ Rev.\ A {\bf 73}, 022309 (2006).

\bibitem{maarten1}
 	M.\ Van den Nest, A.\ Miyake, W.\ D{\"u}r, and
 	H.J.\ Briegel, Phys.\ Rev.\ Lett.\ {\bf 97}, 150504 (2006).
		
\bibitem{InnsbruckLong}
	M.\ van den Nest, W.\ D\"ur, A.\ Miyake, and H.J.\ Briegel,
	quant-ph/0702116.

\bibitem{FCS}
	M.\ Fannes, B.\ Nachtergaele, and R.F.\ Werner,
	Commun.\ Math.\ Phys.\ {\bf 144}, 443 (1992);
	Y.S.\ \"Ostlund and S.\ Rommer, 
	Phys.\ Rev.\ Lett.\ {\bf 75}, 3537 (1995);
	U.\ Schollw{\"o}ck, Rev.\ Mod.\ Phys.\ {\bf 77}, 259 (2005);
	D.\ Perez-Garcia, F.\ Verstraete, M.M.\ Wolf, and
	J.I.\ Cirac, quant-ph/0608197; J.\ Eisert,
	Phys.\ Rev.\ Lett.\ {\bf 97}, 260501 (2006).

\bibitem{david}
	D.\ Perez-Garcia, F.\ Verstraete, M.M.\ Wolf, and J.I.\ Cirac,
	Quant.\ Inf.\ Comp.\ {\bf 7}, 401 (2007).
	
\bibitem{PEPS}
	F.\ Verstraete and J.I.\ Cirac,
	cond-mat/0407066; 
	S.\ Richter (PhD thesis, Osnabr{\"u}ck, 1994), 
	supervised by R.F.\ Werner;
	F.\ Verstraete, M.M.\ Wolf, D.\ Perez-Garcia, 
	J.I.\ Cirac, Phys.\ Rev.\ Lett.\ {\bf 96}, 220601 (2006).

\bibitem{LE}
	 M.\ Popp, F.\ Verstraete, M.A.\ Martin-Delgado, and
	 J.I.\ Cirac, 
	 Phys.\ Rev.\ A {\bf 71}, 042306 (2005).

\bibitem{QCA}
	 B.\ Schumacher and R.F.\ Werner,
	 quant-ph/0405174.

\bibitem{dLevel}
	D.L.\ Zhou, B.\ Zeng, Z.\ Xu, and C.P.\ Sun,
	Phys.\ Rev.\ A {\bf 68}, 062303 (2003).
	
\bibitem{nielsenChuang}
	M.A.\ Nielsen and I.L.\ Chuang,
	{\it Quantum computation and quantum
	information} (Cambridge University Press,
	Cambridge, 2000); J.\ Eisert and M.M.\
	Wolf, {\it Quantum computing}, in 
	{\it Handbook of nature-inspired and innovative 	
	computing} (Springer, New York, 2006).

\bibitem{adiabatic}
	D.\ Aharonov, W.\ van Dam, J.\ Kempe, Z.\ Landau, 
	S.\ Lloyd, and 
	O.\ Regev, quant-ph/0405098.
      
\bibitem{bravyi:toric-mbc}
	S.\ Bravyi and R.\ Raussendorf, quant-ph/0610162.
	
\bibitem{verstraeteDiverging}
	F.\ Verstraete, M.A.\ Mart\'in-Delgado, and J.I.\ Cirac,
	Phys.\ Rev.\ Lett.\ {\bf 92}, 087201 (2004).
	
\bibitem{graphical}
	P.\ Cvitanovic, Phys.\ Rev.\ D {\bf 14}, 1536 (1976);
	R.B.\ Griffiths, S.\ Wu, L.\ Yu, and S.C.\ Cohen,
	Phys.\ Rev.\  A {\bf 73}, 052309 (2006).

\bibitem{Weighted}        
         W.\ D{\"u}r, L.\ Hartmann, M.\ Hein, M.\ Lewenstein, and H.J.\ 
         Briegel, Phys.\ Rev.\ Lett.\ {\bf 94}, 097203 (2005);
          S.\ Anders, M.B.\ Plenio, W.\ D{\"u}r, 
          F.\ Verstraete, and H.-J. Briegel,
          ibid.\ {\bf 97}, 107206 (2006).

\bibitem{kitaev:toriccodes}
	A.Y.\ Kitaev, Ann.\ Phys.\ \textbf{303}, 2 (2003).

\bibitem{frank:PEPS}
	F.\ Verstraete, M.M.\ Wolf, D.\ Perez-Garcia, and J.I.\ Cirac,
	Phys.\ Rev.\ Lett.\ \textbf{96}, 220601 (2006).
	
	
\bibitem{Non}
        J.\ Eisert, K.\ Jacobs, P.\ Papadopoulos, and
        M.B.\ Plenio, Phys.\ Rev.\ A {\bf 62}, 052317 (2000);
        D.\ Collins, N.\ Linden, and S.\ Popescu,
        ibid.\ {\bf 64}, 032302 (2001); 
        D.\ Gottesman, {\it The Heisenberg Representation of 
        Quantum Computers},
	in S.P.\ Corney  et.\ al.\ Eds., Proc.\ XXII Int.\ Coll.\ 
	Group Theor.\ Meth.\ Phys.
	(International Press, Cambridge, 1999);
	J.I.\ Cirac, W.\ D{\"u}r, B.\ Kraus, and M.\ Lewenstein,
	Phys.\ Rev.\ Lett.\ {\bf 86}, 544 (2001).

\bibitem{Priv}
	F.\ Verstraete, private communication.

\bibitem{Percolation}	
	K.\ Kieling, T.\ Rudolph, and J.\ Eisert,
	quant-ph/0611140.
	
\bibitem{Grimmett}
    	G.\ Grimmett, {\it Percolation} (Springer, Berlin, 1999).
	  	 
\bibitem{Walgate}
	J.\ Walgate, A.J.\ Short, L.\ Hardy, and V.\ Vedral,
	Phys.\ Rev.\ Lett.\ {\bf 85}, 4972 (2000).
 
\bibitem{Linear}
	J.\ Eisert, 	
	Phys.\ Rev.\ Lett.\ {\bf 95}, 040502 (2005).	

\end{thebibliography}
\end{document}